\documentclass[aps,superscriptaddress,floatfix]{revtex4}
\usepackage{amsmath}
\usepackage{amsbsy}
\usepackage{amssymb}
\usepackage[dvipdfmx]{graphicx}
\usepackage{epsfig}
\usepackage{color}
\usepackage[normalem]{ulem}
\usepackage{subfigure}
\usepackage{verbatim}
\usepackage{psfrag}
\usepackage{mathrsfs} 
 \usepackage{siunitx} 

\renewcommand{\Re}{\operatorname{Re}}

\newcommand{\K}{\widetilde{K}}
\newcommand{\m}{\widetilde{m}}

\renewcommand{\L}{\operatorname{\mathscr{L}}}
\newcommand{\N}{\operatorname{\mathscr{N}}}
\renewcommand\O[1]{\mathrm{O}\left (#1\right )}

\renewcommand{\d}{\mathrm{d}}

\newcommand{\cc}{\mathrm{c.c.}}

\newcommand{\Psit}{\widetilde{\Psi}}

\def\bse{\begin{subequations}}
\def\ese{\end{subequations}}

\begin{document}
\title{Onset of synchronization in networks of second-order Kuramoto
oscillators with delayed coupling: Exact results and application to phase-locked loops}
\author{David M\'etivier}
\email{metivier@lanl.gov}
\thanks{These authors contributed equally to the work.}
\affiliation{CNLS \& T-4 of Los Alamos National Laboratory, NM 87544, USA}
\author{Lucas Wetzel}
\email{lwetzel@pks.mpg.de}
\thanks{These authors contributed equally to the work.}
\affiliation{\mbox{Max Planck Institute for the Physics of Complex Systems,
N\"{o}thnitzer Stra\ss e 38,
D-01187 Dresden, Germany}}
\author{Shamik Gupta}
\email{shamikg1@gmail.com}
\affiliation{\mbox{Department of Physics, Ramakrishna Mission Vivekananda
Educational and Research Institute, Belur Math, Howrah 711202, India}}
\affiliation{Regular Associate, Quantitative Life Sciences Section, \\ICTP - The Abdus Salam International Centre for Theoretical Physics, Strada Costiera 11, 34151 Trieste, Italy}
\begin{abstract}
We consider the inertial Kuramoto model of $N$ globally coupled oscillators characterized by both their phase and angular velocity, in which there is a time delay in the interaction between the oscillators. 
Besides the academic interest, we show that the model can be related to a network of phase-locked loops widely used in electronic circuits for generating a stable frequency at multiples of an input frequency. 
%
We study the model for a generic choice of the natural frequency distribution of the oscillators, to elucidate how a synchronized phase bifurcates from an incoherent phase as the coupling constant between the oscillators is tuned. 
We show that in contrast to the case with no delay, here the system in the stationary state may exhibit either a subcritical or a supercritical bifurcation between a synchronized and an incoherent phase, which is dictated by the value of the delay present in the interaction and the precise value of inertia of the oscillators. 
Our theoretical analysis, performed in the limit $N \to \infty$, is based on an unstable manifold expansion in the vicinity of the bifurcation, which we apply to the kinetic equation satisfied by the single-oscillator distribution function. 
We check our results by performing direct numerical integration of the dynamics for large $N$, and highlight the subtleties
arising from having a finite number of oscillators. 
\end{abstract}
\date{\today}
\maketitle

\section{Introduction}
\label{sec:introduction}

\subsection{The model}

The Kuramoto model with inertia is representative of complex many-body
dynamics involving a set of rotors characterized by their phases and
angular velocities that are coupled all-to-all through the sine of their
phase differences. Specifically, the dynamics for a system of $N$ rotors is given by a set of $2N$ coupled first-order differential equations of the form~\cite{Tanaka:1997,Acebron:1998,Acebron:2000}
\begin{eqnarray}
&&\dot{\theta}_i(t)=v_i(t), \nonumber \\
\label{eq:Kuramoto-inertia-delay-0} \\
&&m\dot{v}_{i}(t)=-\gamma
v_{i}(t)+\gamma\omega_{i}+\frac{K}{N}\sum_{j=1}^{N}\sin(\theta_{j}(t)-\theta_{i}(t))\nonumber,
\end{eqnarray}
where the dot denotes derivative with respect to time, $\theta_i \in [0,2\pi)$ and $v_i$ are the phase and the angular
velocity of the $i$-th rotor, respectively, whose moment of inertia is
$m>0$. Here, $\gamma>0$ is the damping constant, $K>0$ is the coupling constant,
while $\omega_i \in [-\infty,\infty]$ is the natural frequency of the
$i$-th rotor. The frequencies $\{\omega_i\}_{1\le i \le N}$ constitute a
set of independent and quenched disordered random variables distributed
according to a given distribution $G(\omega)$, normalized as
$\int_{-\infty}^\infty {\rm d}\omega~G(\omega)=1$ and with finite mean $\omega_0$. 
During the analysis we will also use the centered distribution $g(\omega) \equiv G(\omega+\omega_0)$.
In the limit of overdamping, $\gamma/m \to \infty$, the rotors are effectively characterized by their phases alone and are therefore quite rightly referred to as oscillators \footnote{In this work, we use the terms ``oscillators" and ``rotors" interchangeably.}. In this limit, the dynamics~(\ref{eq:Kuramoto-inertia-delay-0}) becomes that of the Kuramoto model~\cite{Kuramoto:1984,Strogatz:2000,Acebron:2005,Gupta-Campa-Ruffo:2014-2,Rodrigues:2016,Gherardini:2018,Gupta:2018}, which over the years has emerged as a paradigmatic minimal framework to study spontaneous collective synchronization in a group of coupled limit-cycle oscillators, such as that observed in groups of fireflies flashing on and off in
unison~\cite{Buck:1988}, in cardiac pacemaker cells~\cite{Peskin:1975}, in Josephson
junction arrays~\cite{Benz:1991}, 
in electrochemical~\cite{Kiss:2002} and electronic~\cite{Temirbayev:2012} oscillators, etc. The governing equations of the Kuramoto model are $N$ coupled first-order differential equations of the form
\begin{equation}
\gamma\dot{\theta}_i(t)=\gamma \omega_i+\frac{K}{N}\sum_{j=1}^N \sin(\theta_j(t)-\theta_i(t)). \label{eq:Kuramoto-eom}   
\end{equation}
The mean-field nature of either the dynamics~(\ref{eq:Kuramoto-eom}) or the dynamics~(\ref{eq:Kuramoto-inertia-delay-0}) becomes evident on defining the so-called Kuramoto order parameter $R(t)$ and the global phase $\Phi(t)$, as~\cite{Kuramoto:1984}
\begin{equation}
R(t)e^{i\Phi(t)}\equiv \frac{\sum_{j=1}^N e^{i\theta_j(t)}}{N};~~R,\Phi
\in \mathbb{R},~0 \le R \le 1,~\Phi \in [0,2\pi),  
\label{eq:kuramoto-r}
\end{equation}
with $0<R<1$ characterizing a synchronized phase, and $R=0$ an incoherent phase. 
In terms of $R(t)$, the dynamics~(\ref{eq:Kuramoto-inertia-delay-0}) may be rewritten as
\begin{eqnarray}
&&\dot{\theta}_i(t)=v_i(t), \nonumber \\
\label{eq:Kuramoto-inertia-delay-0-rt} \\
&&m\dot{v}_{i}(t)=-\gamma
v_{i}(t)+\gamma\omega_{i}+KR(t)\sin(\Phi(t)-\theta_{i}(t))\nonumber,
\end{eqnarray}
which shows that the evolution of the dynamical variables at time $t$ is governed by the value of the mean-field $R(t)e^{i\Phi(t)}$ set up collectively at time $t$ by all the oscillators.

Both the models~(\ref{eq:Kuramoto-inertia-delay-0}) and
(\ref{eq:Kuramoto-eom}) have been extensively studied in the past and a
host of results have been derived with regard to the parameter regimes allowing for the emergence of a synchronized stationary state (see Ref.~\cite{Gupta:2018}
for a recent overview). For example, consider a $G(\omega)$ that is unimodal, namely, it is symmetric about its mean $\omega_0$, and decreases monotonically and
continuously to zero with increasing $|\omega - \omega_0|$. In this case, it is known that in the stationary state of the dynamics (\ref{eq:Kuramoto-eom}), the system for a given choice of $G(\omega)$ may exist in either a synchronized or an incoherent phase depending on whether the coupling $K$
is respectively above or below a critical value $K_c=2/(\pi G(\omega_0))$; on tuning $K$ across $K_c$ from high to low
values, one observes a \textit{continuous} phase transition in $R_{\rm st}$, the
stationary value of $R(t)$. 
Namely, $R_{\rm st}$ decreases continuously from the value of unity, achieved as $K \to \infty$, to the value zero at $K=K_c$ and remains zero at smaller $K$ values.
One may interpret the transition as
the case of a supercritical bifurcation, in which on tuning $K$, a synchronized phase bifurcates from the incoherent phase at $K=K_c$. In particular, a small change of $K$ across $K_c$ results in only a small change in the value of $R_{\rm st}\propto \sqrt{K-K_c}$ close to and above $K_c$~\cite{Kuramoto:1984,Crawford:1995}. For the same choice of a unimodal $G(\omega)$, the inertial dynamics~(\ref{eq:Kuramoto-inertia-delay-0}) on the other hand show a \textit{discontinuous}  phase transition between synchronized
and incoherent phase, where $R_{\rm st}$ exhibits an abrupt and big change
from zero to a non-zero value on changing $K$ by a small amount across the
phase transition point~\cite{Olmi:2014,Barre:2016}. Here, the bifurcation of
the synchronized from the incoherent phase is said to be subcritical and
leads to hysteresis~\cite{Strogatz-book}. Thus, presence of
inertia is rather drastic in that it changes completely the nature of
the bifurcation and hence of the underlying stationary state.

In this work, we study for the first time the effect of a delay in the
interaction between the oscillators within the framework of
dynamics~(\ref{eq:Kuramoto-inertia-delay-0}). The dynamical equations of this modified model are given by 
\begin{eqnarray}
\dot{\theta}_i(t)&=&v_i(t), \nonumber \\
\label{eq:Kuramoto-inertia-delay-1-rt} \\
m\dot{v}_{i}(t)&=&-\gamma
v_{i}(t)+\gamma\omega_{i}+KR(t-\tau)\sin(\Phi(t-\tau)-\theta_{i}(t)-\alpha)\nonumber,
\end{eqnarray}
thereby modeling the time-evolution which is governed by the value of the mean field at an earlier instant $t-\tau$, where $\tau>0$ is the time delay in the interaction between the oscillators. Here, $\alpha \in (-\pi/2,\pi/2)$ is the so-called phase frustration parameter, an additional dynamical parameter that is known to affect significantly the behavior of the Kuramoto model~\cite{Sakaguchi:1986}. In the overdamped limit, the model~(\ref{eq:Kuramoto-inertia-delay-1-rt}) reduces to 
\begin{equation}
\gamma \dot{\theta}_{i}(t)=\gamma\omega_{i}+KR(t-\tau)\sin(\Phi(t-\tau)-\theta_{i}(t)-\alpha),
\label{eq:Kuramoto-inertia-delay-1-rt-no-m}
\end{equation}
which in presence of additional Gaussian, white noise has been addressed
in Ref.~\cite{Yeung-Strogatz:1999}. Note that for the dynamics in
(\ref{eq:Kuramoto-inertia-delay-1-rt-no-m}), the parameter $\gamma$
may be scaled out by a redefinition of time, so that the relevant
dynamical parameters are $K,\tau$ and $\alpha$. In a recent
work~\cite{Metivier:2019}, two of the present
authors have investigated the
dynamics~(\ref{eq:Kuramoto-inertia-delay-1-rt-no-m}), deriving for generic
$G(\omega)$ and as a function
of the delay exact results for the stability boundary $K_c(\tau)$
between the incoherent and the synchronized phase and the nature in which the
latter bifurcates from the former at the phase transition point.
Note that unlike~(\ref{eq:Kuramoto-inertia-delay-0}), the
dynamics~(\ref{eq:Kuramoto-inertia-delay-1-rt}) is not invariant under the transformation $\theta_j(t)
\to \theta_j(t) - \omega_0 t,~\omega_j \to \omega_j - \omega_0~\forall~j$ that views the dynamics in a frame rotating uniformly at
frequency $\omega_0$ with respect to an inertial frame. From
Eq.~(\ref{eq:Kuramoto-inertia-delay-1-rt}), it is clear that viewing the dynamics in such a frame is equivalent to replacing
$\alpha$ with $\alpha'\equiv
\alpha+\omega_0 \tau$.
Our results imply that for a given choice of $G(\omega)$, the nature of
transition (continuous versus discontinuous) between the synchronized and incoherent phases depends
explicitly on the value of $\tau$. 

In view of the aforementioned developments, it is evidently of interest to investigate
the effects of inertia on the time-delayed model and thus embark on a
detailed analysis of the dynamics~(\ref{eq:Kuramoto-inertia-delay-1-rt}). Since even without delay, inertia is known to have nontrivial and interesting consequences as mentioned above, we may already anticipate that an interplay of the influence of
delay and inertia may result in an even richer stationary state for the
dynamics~(\ref{eq:Kuramoto-inertia-delay-1-rt}) vis-\`{a}-vis for dynamics~(\ref{eq:Kuramoto-inertia-delay-1-rt-no-m}). Remarkably, the dynamics in Eq.~(\ref{eq:Kuramoto-inertia-delay-1-rt}), far from being just a model of academic interest, emerge naturally in the context of mutually coupled phase-locked loops, as we now demonstrate. 

\subsection{Relation to a network of phase-locked loops}

\begin{figure}[!ht]
\centering
\includegraphics[width=10cm]{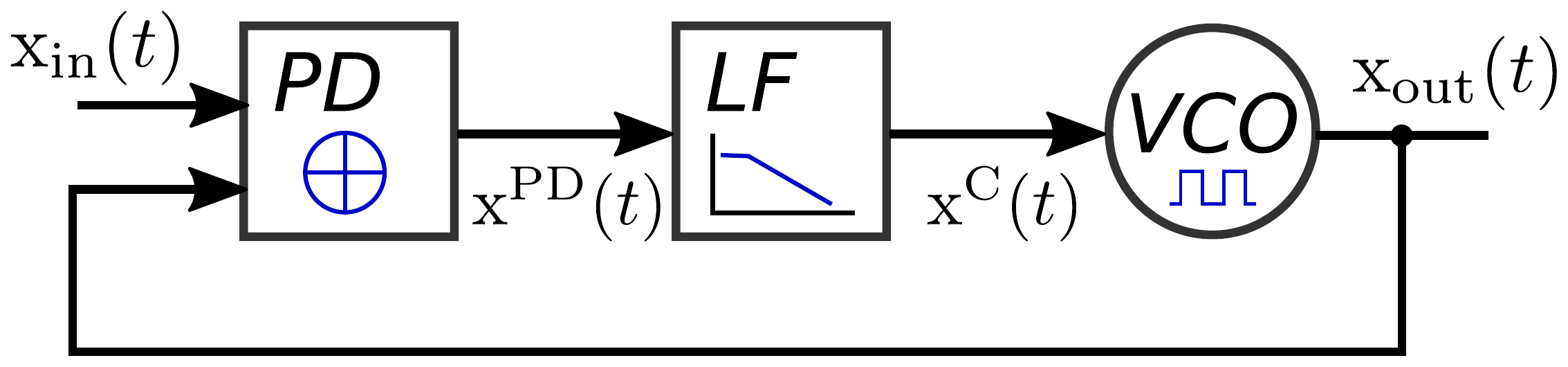}
\caption{Schematic diagram of a phase-locked loop (PLL). The arrows
denote the direction of flow of signals in the loop.}
\label{fig:pll}
\end{figure}

A phase-locked loop (PLL) is an electronic component designed to generate an output signal that has a constant phase relation (and is thus locked) to the phase of its input reference. Fig.~\ref{fig:pll} shows a schematic PLL architecture consisting of a phase detector (PD), a loop filter (LF), and a voltage-controlled oscillator (VCO) acting as a variable-frequency oscillator, all connected in a feedback loop.
The phase-detector output $x^{\rm PD}(t)$ represents the phase relations of the periodic output signal $x_{\rm out}(t)$, generated by the VCO,
with the phase of the periodic input-signal $x_{\rm in}(t)$.
The loop-filtered phase-detector output yields the control signal $x^{\rm C}(t)$ that controls the instantaneous frequency of the VCO so that its corresponding output approaches the phase and frequency of the input signal. 
The latter property enables a PLL to track an input frequency, or, to generate a frequency that is a multiple of the input frequency. 
PLL's find wide use in electronic applications as an effective device to, e.g., recover a signal from a noisy communication channel, generate a stable frequency at multiples of an input frequency, and to distribute
a quartz reference clock signal via a clocktree architecture.

Let us now consider the setup of $N \ge 2$ mutually delay-coupled PLL's occupying the nodes of a network, in which the input signals for a given PLL are constituted by the delayed output received from other PLL's~\cite{Pollakis:2014,Joerg:2015,Wetzel:2017}. 
The delay could be due to transmission signaling-times, and is accounted for in the following by a discrete delay-time $\tau$.
We consider the LF to ideally damp the high-frequency components of the PD signal.
Consider the output signal of the $i$-th PLL, $i=1,2,\ldots,N$, in the network
\begin{equation}
x_{i}(t)=\operatorname{sig}\left(\theta_{i}(t)\right),
\label{eq:xk}
\end{equation}
where $\theta_{i}(t)$ denotes the phase, and $\operatorname{sig}(\theta_i(t))$ is a $2\pi$-periodic function with amplitude one.
Depending on the type of PLL, i.e., analog or digital, the output signal may be sinusoidal or a rectangular function, respectively.
The VCO is operated such that its output frequency $\dot{\theta}_i(t)$
depends linearly on the control signal $x_i^{\rm C}(t)$: 
\begin{equation}
\dot{\theta}_i(t)=\omega_{i,0}^{\rm VCO}+K_i^{\rm VCO}\,x_i^{\rm C}(t),
\label{eq:VCOdyn}
\end{equation}
where $\omega_{i,0}^{\rm VCO}$ denotes the natural frequency, and $K_i^{\rm VCO}$ the VCO input sensitivity. 
The control signal is the output of the loop-filter:
\begin{equation}
x_i^{\rm C}(t)=\int_0^{\infty}\,\textrm{d}u~p(u)\,x_i^{\rm PD}(t-u),
\label{eq:LFint}
\end{equation}
where $x_i^{\rm PD}(t)$ denotes the phase-detector signal, and $p(u)$ is the impulse response of the filter.
Considering first-order loop-filters, i.e., $p(u;1,b)$ being the
$\Gamma$-distribution with shape parameter $a=1$ and scale parameter
$b$, the above integral equation can be rewritten by using Laplace
transforms~\cite{Mancini:2003,Wetzel:2017}, yielding
\begin{equation}
\dot{x}_i^{\rm C}(t)=\omega_c\left[x_i^{\rm PD}(t)-x_i^{\rm C}(t)\right],
\label{eq:LFdiff}
\end{equation}
where $\omega_c$ denotes the cut-off frequency of the first-order low-pass filter.
The initial state of the filter is given by $x_i^{\rm C}(0)=(\dot{\theta}_i(0)-\omega_{i,0}^{\rm VCO})/K_i^{\rm VCO}$.
The phase-detector signal depends on the type of PLL:
\begin{equation}
x_i^{\rm PD}(t)=C+\frac{1}{2\,n(i)}\sum\limits_{j=1}^N\,c_{ij}h\left[\theta_j(t-\tau)-\theta_i(t) \right],
\label{eq:PDgeneral}
\end{equation}
where $C$ is a PLL type specific offset ($C=1/2$ for XOR PD's, while $C=0$ for multiplier PD's), $c_{ij}=\{0,1\}$ are the components of the adjacency matrix, with the value $1$ (respectively, $0$) denoting whether PLL units $i$ and $j$ are coupled (respectively, uncoupled), $n(i)\equiv \sum_j
c_{ij}$ the total number of units coupled with unit $i$, $h(x)$ is a
$2\pi$-periodic coupling function, and we assumed the high-frequency components to be filtered ideally by the LF~\cite{Pollakis:2014}.
Equations~(\ref{eq:VCOdyn})-(\ref{eq:PDgeneral}) combined together yield a
second-order phase model with delayed-coupling:
\begin{equation}
\frac{1}{\omega_c}\ddot{\theta}_i(t)+\dot{\theta_{i}}(t)=\omega_{i}+\frac{\widetilde{K}_i}{n(i)}\sum_{j=1}^{N}c_{ij}h(\theta_{j}(t-\tau)-\theta_{i}(t))\,.
\label{eq:2ndOrderKuramoto}
\end{equation}
where we have defined $\omega_{i}\equiv\omega_{i,0}^{\rm VCO}+C\,K_i^{\rm VCO}$, and $\widetilde{K}_i\equiv K_i^{\rm VCO}/2$. 
The $2\pi$-periodic coupling function $h$ depends on the type of the PD and the corresponding input signals.
Here we consider the case of a cosine coupling function, $h(x)=\cos(x)$,
for analog PLL's and multiplier phase-detectors and a triangular coupling
function, and $h(x)=\Delta(x)$ for digital PLL's with XOR phase-detectors.
In the latter case the coupling-function can be approximated as $h(x)=-8/\pi^2\cos(x)$. 
The case of a d-flip flop \footnote{Note that in most state-of-the-art
electronic systems where synchronization is achieved through entrainment
by a reference clock, the phase detector is a \textit{flip-flip}
phase-frequency detector, contrary to the XOR component used for the
phase detector of the digital PLL's discussed in this work.} phase
detector for digital PLLs, which has a linear coupling-function, will not be
considered in this work.
Given these cases, we will use a sinusoidal coupling-function with a
phase frustration parameter $\alpha\,\in\,[-\pi/2, \pi/2]$, that is,
with $h(x)=\sin(x-\alpha)$, which represents both of the cases mentioned
above. We will also specialize to the case when every PLL unit is
coupled to every other, implying $c_{ij}=1~\forall~i,j=1,2,\ldots,N$ and $n(i)=N$. 
Comparing Eqs.~(\ref{eq:2ndOrderKuramoto}) and~(\ref{eq:Kuramoto-inertia-delay-1-rt}) leads to the correspondence $m=\omega_c^{-1}=b,~\gamma=1$, as well as $K=\widetilde{K}_i,~\alpha=-\pi/2$ for
the analog PLL case, and $\alpha=\pi/2,~K=8\widetilde{K}_i/\pi^2$ for the digital PLL approximation.

Before moving on to an analysis of the
dynamics~(\ref{eq:Kuramoto-inertia-delay-1-rt}), it is pertinent that we
give here a summary of our results obtained in this paper and the
techniques employed in achieving them. We here obtain exact analytical relations for the critical point $K_c(\tau)$ beyond which the incoherent phase of the dynamics~(\ref{eq:Kuramoto-inertia-delay-1-rt}) becomes unstable, and furthermore, the nature in which the synchronized phase bifurcates from the incoherent
phase as $K$ is increased beyond $K_c(\tau)$. An illustration of our
results for a unimodal Lorentzian distribution is shown in
Fig.~\ref{fig:del:strog_m} for two representative values of the
inertia, which
displays both $K_c(\tau)$ and $s(\tau)$ whose sign determines 
the nature of the bifurcation of the order parameter $R$, Eq.~\eqref{eq:kuramoto-r}: a positive (respectively, a negative) sign
implies a subcritical bifurcation and hence, a discontinuous transition
(respective, a supercritical bifurcation and hence a continuous
transition). As may be seen from the Fig.~\ref{fig:del:strog_m}, $K_c$ and $s$ both have
an essential dependence on $\tau$ and $m$, while our analysis (see Eq.~(\ref{eq:del:c3eqbis}))
suggests that the effects of changing $\tau$ at a fixed $\alpha$ are the same as those
from changing $\tau$ at a
fixed $\alpha$ keeping $\alpha + \omega_0\tau$ constant.

We now summarize our method of analysis in obtaining the aforementioned results. We start off with considering the
dynamics~(\ref{eq:Kuramoto-inertia-delay-1-rt}) in the limit $N \to
\infty$, when it may be effectively characterized by a
single-oscillator probability density $F(\theta,v,\omega,t)$, which
gives at time $t$ and for each $\omega$ the fraction of oscillators with
phase $\theta$ and angular velocity $v$. The time evolution of
$F(\theta,v,\omega,t)$ follows a kinetic equation, of which the
incoherent state $f^0(\theta,v,\omega)$ (associated with $R_{\rm st}=0$) represents a stationary solution. We
rewrite the kinetic equation in the form of a delay differential
equation (DDE)~\cite{Hale:1963,Hale:1993} for perturbations $f_t(\varphi) \equiv F(\theta,v,\omega,t+\phi);~-\tau \le \phi < 0$ around
$f^0(\theta,v,\omega)$. The
DDE involves a linear evolution operator $\mathscr{D}$ and a nonlinear
one, $\mathscr{F}$. We obtain the eigenvalues and the eigenvectors of $\mathscr{D}$ and
of the corresponding adjoint operator $\mathscr{D}^\dagger$. As is well
known~\cite{Strogatz-book}, the knowledge of the eigenvalues allows to
locate the critical value $K_c$ of the coupling $K$ above which the
incoherent state $f^0(\theta,v,\omega)$ becomes linearly unstable. We then build for $K>K_c$ the unstable manifold expansion of the perturbation $f_t(\phi)$ along the two complex conjugated eigenvectors associated with the instability.
Using a convenient Fourier expansion of the relevant quantities and
working at $K$ slightly greater
than $K_c$, we thus obtain the amplitude dynamics describing the evolution of
perturbations $f_t(\varphi)$ in the regime of weak linear instability,
$K\to K_c^+$. The nature of the amplitude dynamics at once dictates the
nature of bifurcation occurring as soon as $K$ is increased beyond
$K_c$: The amplitude dynamics has a leading linear term and a nonlinear
(cubic) term, and as is well known from the theory of bifurcation
\cite{Strogatz-book}, the sign of the real part of this cubic term (denoted $s$ in Fig.~\ref{fig:del:strog_m}) dictates the nature of
the bifurcation, with positive and negative signs leading respectively
to subcritical and supercritical bifurcation.

The paper is organized as follows. Section~\ref{sec:Kuramoto-analysis}
forms the core of the paper, in which we derive our main results,
Eqs.~(\ref{eq:del:Lambda}) and (\ref{eq:del:c3eqbis}).
We illustrate our analytical results with the representative example of
a unimodal Lorentzian distribution. In Section~\ref{sec:numerics}, we
make a detailed comparison of our analytical results obtained in the
limit $N \to \infty$ with numerical results for finite $N$ obtained by
performing numerical integration of the equations of
motion. Here, in particular, we
discuss the subtleties involved in making such a comparison whose origin
may be traced to finite-size effects prevalent for finite $N$. The paper ends with conclusions. 

\section{Exact analysis in the limit $N \to \infty$}
\label{sec:Kuramoto-analysis}

We now turn to a derivation of our results for the system~(\ref{eq:Kuramoto-inertia-delay-1-rt}). To simplify matters, we work in the rotating frame $\theta_j(t)
\to \theta_j(t) - \omega_0 t,~\omega_j \to \omega_j - \omega_0~\forall~j$, so that the distribution $G(\omega) \to g(\omega)$ is now centered in 0.
Moreover, consider the system in the limit $N \to \infty$, when
the dynamics may be effectively characterized in terms of the
single-oscillator probability density $F(\theta,v,\omega,t)$ defined
above. This density is $2\pi$-periodic
in $\theta$, and obeys the
normalization 
\begin{equation}
\int_0^{2\pi}{\rm d}\theta \int_{-\infty}^\infty {\rm
d}v~F(\theta,v,\omega,t)=g(\omega)~\forall~\omega,t.
\end{equation}
The time evolution of
$F(t)\equiv F(\theta,v,\omega,t)$ may be derived by following the procedure given in Ref.~\cite{Gupta-Campa-Ruffo:2014-2}. 
One obtains the evolution equation
\begin{eqnarray}
&\dfrac{\partial F}{\partial t}(t)+ v\dfrac{\partial F}{\partial \omega}(t)+\dfrac{K}{2i m} \left
(R_1[F](t-\tau)  e^{-i(\theta+\alpha+\omega_0\tau})-R_{-1}[F](t-\tau)
e^{i(\theta+\alpha+\omega_0\tau})\right )\dfrac{\partial F}{\partial v} (t)-\dfrac{\gamma}{m}\dfrac{\partial }{\partial v}\left
((v-\omega)F(t)\right)=0,\nonumber\\
\label{eq:del:pde}
\end{eqnarray}
where we have defined as functionals of $F$ the quantity
\begin{equation}
R_k[F]\equiv \int_0^{2\pi} {\rm d}\theta \int_{-\infty}^\infty {\rm d}v \int
{\rm d}\omega~e^{ik\theta}F(\theta,v,\omega,t);~~k=0,\pm 1,\pm 2,\ldots.
\end{equation}
In particular, $R_{1}$ coincides with the $N \to \infty$ limit of the
Kuramoto complex order parameter $R(t)e^{i\Phi(t)}$ in Eq.~(\ref{eq:kuramoto-r}), and hence $|R_{-1}|=|R_1|=R$.

From Eq.~(\ref{eq:del:pde}), one may check that the incoherent state
\begin{equation}
f^0(\theta,v,\omega)=g(\omega) \frac{\delta(v-\omega)}{2\pi}
\end{equation}
solves the equation in the stationary state and thus represents an
incoherent stationary state. To examine how in the stationary state the incoherent stable becomes unstable as $K$ is tuned above a critical value $K_c$, we employ an unstable
manifold expansion of perturbations about the incoherent state in the
vicinity of the bifurcation. 
To perform the analysis, we write $F=f^0+f$, with $f$ being the
perturbation. Next, we note that the time evolution of the function
$F(t)$ according to a nonlinear operator with delay $M[F(t)]$ (obtained
from Eq.~(\ref{eq:del:pde})) can be rewritten in term of a delay variable $\varphi$ such
that the time-evolution operator is given by
\begin{equation}
(\mathscr{A} F_t)(\varphi)=
\begin{cases}
\dfrac{\d}{\d\varphi}F_t(\varphi),\quad -\tau\leq\varphi\leq 0,\\
M[F_t],\qquad \varphi=0,
\end{cases}
\label{eq:D+N}
\end{equation}
with $F_t(\phi) \equiv F(t+\phi)$. 
Employing the expansion $F=f^0+f$, we define the linear and nonlinear operators
$\mathscr{D}$ and $\mathscr{F}$, according to
\begin{equation}
(\mathscr{A} f_t)(\varphi)=(\mathscr{D}f_t+\mathscr{F}[f_t] )(\varphi)=\begin{cases}
\dfrac{\d}{\d\varphi}f_t(\varphi)\\
\mathscr{L}f_t(\varphi)
\end{cases}
+
\begin{cases}
0,\qquad~~~~ -\tau\leq\varphi< 0,\\
\mathscr{N}[f_t],\quad \varphi=0.
\end{cases}
\label{eq:del:delay_operator}
\end{equation}
We decompose the linear operator into two parts,
$\L=\operatorname{L}+\operatorname{\mathcal{R}}$, namely, a part $\operatorname{L}$ that
does not contain any delay term and a part $\operatorname{\mathcal{R}}$ that has
all the delay terms.
Rewriting Eq.~(\ref{eq:del:pde}) according to the above formalism yields
\begin{equation}
\dfrac{\partial f_t}{\partial t}=\mathscr{D} f_t+\mathscr{F} [f_t],
\label{eq:del:VKdec-0}
\end{equation}
with 
\begin{eqnarray}
&&        \operatorname{L} f=-v\partial_\theta f+\dfrac{1}{\m}\partial_v\left ((v-\omega)f\right ),\\
&&        \operatorname{\mathcal{R}} f=-\dfrac{\K}{2i \m}\left (R_1[f]
e^{-i\theta}e^{-i(\alpha+\omega_0\tau)}-R_{-1}[f] e^{i\theta}e^{i(\alpha+\omega_0\tau)}\right )\partial_v f^0,
\\
&&\N [f_t]=-\dfrac{\K}{2i \m}\left (R_1[f_t](-\tau)
e^{-i\theta}e^{-i(\alpha+\omega_0\tau)}-R_{-1}[f_t](-\tau) e^{i\theta}e^{i(\alpha+\omega_0\tau)}\right )\partial_v f(0),\label{eq:N}
\end{eqnarray}
where we use the shorthand $\partial_v \equiv \partial /\partial v$ for
derivatives and use from now on the transformation $\m = m/\gamma$ and
$\K = K/\gamma$. 

In the functional space of delayed functions, there is no $\mathcal{L}_2$ canonical
inner product. However,
Ref.~\cite{Hale:1963} defines a bilinear form acting as the inner product on this space. 
In our problem with a discrete delay, the scalar product is
\begin{eqnarray}
&&(q,p)_\tau\equiv
(q(0),p(0))
+\int_{-\tau}^0 {\mathrm d}\xi~ \left(q(\xi+\tau),\operatorname{\mathcal{R}} p(\xi)\right),
\label{eq:del:appli_scalar}
\end{eqnarray}
where $\left (q(0),p(0)\right )$ denotes the usual scalar product on
$\mathcal{L}_2(\mathbb{T}\times\mathbb{R}\times\mathbb{R} )$ (phase,
angular velocity and natural frequency)
\begin{eqnarray}
\left( h,f\right)=\int_{\mathbb{T}\times\mathbb{R}\times\mathbb{R}}h^\ast(\theta,v,\omega) f(\theta,v,\omega)\,\mathrm{d}\omega\,\mathrm{d} v\,\mathrm{d} \theta ,\quad \text{with}\quad q(0)=h(\theta,v,\omega),~p(0)=f(\theta,v,\omega)
\end{eqnarray}
and the integral term contains the delay contribution. 
The adjoint of the linear operator $\mathscr{D}$, obtained by using the
equality
$(q(\varphi),\mathscr{D}p(\varphi))_\tau=(\mathscr{D}^\dagger
q(\varphi),p(\varphi))_\tau$, is defined in the dual space, and is
given by
\begin{equation}
(\mathscr{D}^\dagger q_t)(\vartheta)=\begin{cases}
-\dfrac{\d}{\d\vartheta}q_t(\vartheta),\qquad 0<\vartheta\leq\tau, \\
\mathscr{L}^\dagger q_t(\vartheta),\qquad~~ \vartheta=0.
\end{cases}
\label{eq:adj_D}
\end{equation}
We also decompose $\L^\dagger= \operatorname{L}^\dagger+ \operatorname{\mathcal{R}}^\dagger $, with
\begin{eqnarray}
&&         \operatorname{L}^\dagger q=v\partial_\theta  q -\dfrac{1}{\m}(v-\omega)\partial_v  q,\\
&&        \operatorname{\mathcal{R}}^\dagger  q=\dfrac{\K}{2i \m}\left(
e^{i(\alpha+\omega_0\tau)}e^{-i\theta}R_{1}[q\partial_v f^0]-e^{-i(\alpha+\omega_0\tau)}e^{i\theta}R_{-1}[ q\partial_v f^0] \right).
\end{eqnarray}

Starting with Eq.~(\ref{eq:del:VKdec-0}), the unstable manifold expansion involves a
linear and a weakly nonlinear analysis, and requires combining
two formalisms: i) the one developed in Ref.~\cite{Barre:2016} for the case of the Kuramoto model with inertia but with no delay, ii) the delay formalism
\cite{Hale:1963,Hale:1993,Guo:2013}, as done in
Ref.~\citep{Metivier:2019}. In the following subsections, we go over
one-by-one the various steps, which culminate in our main equation, Eq.~(\ref{eq:A_red}).
~

\subsection{Linear stability analysis of $f^0$}
\label{sec:app:lin}

The linear stanility analysis of the stationary state $f^0$ consists in
solving the eigenvalue problem
\begin{equation}
(\mathscr{D}P)(\varphi)=\lambda P(\varphi)
\label{eq:Deigen}
\end{equation}
for $- \tau\leq\varphi<0$; we get for $\varphi\neq 0$, $P(\varphi)=\Psi e^{\lambda\varphi}$
for arbitrary $\Psi$. 
We expand in a Fourier series in $\theta$, as 
$P(\varphi)=(2\pi)^{-1}\sum_{k=-\infty}^\infty p_k(\varphi) e^{i k\theta}$ and
$\Psi(\theta,\omega)=(2\pi)^{-1}\sum_{k=-\infty}^\infty
\psi_k(\omega)e^{ik\theta}$.
Using Eq.~(\ref{eq:Deigen}) for
$\varphi=0$ and $k=\pm 1$ in the Fourier expansion, we
get 
\begin{equation}
p_1(\varphi)=\psi_1(\omega,v)e^{\lambda \varphi}.
\label{eq:p}
\end{equation}
In the following, we will omit subscripts
while referring to $\psi_1$ and $p_1$. 
For $\varphi=0$, we look for a solution of the eigenvalue problem in the form 
\begin{equation}
\psi = U_0(\omega)\delta(v-\omega) +U_1(\omega)\delta'(v-\omega),
\label{eq:del:varpsi}
\end{equation}
where the Dirac delta function and its derivatives are to be understood
in the distribution sense. Imposing the normalization
$R_{1}[\Psi]=R_{-1}[\Psi]=\int \,\d v\,\d\omega~ \psi=1$, one finds
\begin{eqnarray}
U_0 &=&
\frac{\K}{2\m}e^{i(\alpha+\omega_0\tau)}e^{-\lambda\tau}\frac{g(\omega)}{(\lambda+i\omega)(\lambda
+1/\m +i\omega)}, \\
U_1 &=& \frac{\K}{2i\m}e^{i(\alpha+\omega_0\tau)}e^{-\lambda\tau}\frac{g(\omega)}{\lambda +1/\m +i\omega}.
\end{eqnarray}
Expliciting the normalization condition yields the dispersion relation:
\begin{equation}
\Lambda(\lambda) = 1-\frac{\K}{2\m}e^{i(\alpha+\omega_0\tau)}e^{-\lambda\tau} \int \d\omega \frac{g(\omega)}{(\lambda+i\omega)(\lambda +1/\m +i\omega)} =0.
\label{eq:del:Lambda1}
\end{equation}
We can see that 
$p^\ast(\varphi)$ gives another eigenfunction of $\mathscr{D}$
with eigenvalues $\lambda^\ast$, so that $\Lambda(\lambda)=\Lambda^\ast(\lambda^\ast)=0$.
For $k\neq\pm 1$, one has only a continuous spectrum occupying the imaginary axis.

The adjoint eigenvector has the form $Q(\vartheta)=\Psit
e^{-\lambda^\ast\vartheta}=\widetilde{\psi}e^{i\theta}
e^{-\lambda^\ast\vartheta}$, where $\widetilde{\psi}$ solves 
\begin{equation}
 (\lambda^\ast -iv)\widetilde{\psi} +\frac{1}{\m}(v-\omega)\partial_v
 \widetilde{\psi} = \frac{\K}{2i\m}e^{-i(\alpha+\omega_0\tau)}e^{-\lambda^\star\tau} \int \d\omega \,g(\omega)
 \partial_v \widetilde{\psi}(\omega,\omega).
 \label{eq:del:tildevarpsi}
\end{equation}
Full solution of the above equation is not straightforward to obtain,
but thankfully we just need
to know $\widetilde{\psi}(\omega,\omega)$ and the derivative
$\widetilde{\psi}^{(n)}(\omega)=\partial^{n}_v\widetilde{\psi}(\omega,\omega)$,
which may be obtained by successive differentiation of
Eq.~(\ref{eq:del:tildevarpsi}).

Summarizing, the linear stability analysis yields the dispersion relation 
\begin{equation}
\Lambda(\lambda) \equiv 1-\frac{K}{2m}e^{i(\alpha+\omega_0\tau)}e^{-\lambda\tau} \int \frac{G(\omega+\omega_0)}{(\lambda+i\omega)(\lambda +\gamma/m +i\omega)}\,\d\omega =0,
\label{eq:del:Lambda}
\end{equation}
which has its roots giving the eigenvalues associated with the linear operator
$\mathscr{D}$.
In particular, for $K\geq K_c$, the stationary state $f^0$ becomes unstable, with associated unstable eigenvalues $\lambda$ satisfying $\Re(\lambda)\geq 0$. 
Note that for $K<K_c$, the incoherent state is neutrally stable, i.e., there is no discrete
eigenvalue but only a continuous spectrum; in this case, perturbations
$f^0$ are damped in time via a mechanism similar to the Landau damping \cite{mouhot_landau_2011}.

\subsection{Weakly nonlinear analysis and the unstable manifold expansion}
\label{sec:app:um}

The weakly nonlinear analysis describes the type of bifurcation as $K\to K_c^+$ and hence as $\Re(\lambda)\to 0^+$. 
The analysis involves decomposing the perturbation into a
contribution along the unstable eigenvectors $P(\phi)$,
$P^\ast(\phi)$ associated with the unstable eigenvalues
$\lambda$, $\lambda^\ast$ and a contribution $S_t(\phi)$ in the perpendicular direction, as
\begin{equation}
f_t(\phi)=(A(t)P(\phi)+\cc )+S_t(\phi),
\end{equation}
where $\rm{c.c.}$ stands for complex conjugation, $A(t)=(Q,f_t)_\tau$ is the amplitude of the
unstable mode, and $(Q,S_t)_\tau=0$. Here, we have introduced the eigenvector $Q$ of the adjoint operator $\mathscr{D}^\dagger$ and the scalar product $(\cdot,\cdot)_\tau$. The unstable manifold approach consists in expanding the perpendicular
component $S_t$ in terms of the small amplitude $A$,
$S_t(\phi) \equiv S_t(\theta,v,\omega,\phi)=H[A,A^\ast](\theta,v,\omega,\phi)$ and
computing $H$ perturbatively. We now follow the nonlinear study based on ideas developed in
Refs.~\cite{Barre:2016,Metivier:2019}, and detail our analysis. The
starting point is the expansion 
\begin{equation}
f_t(\varphi)=A(t)P(\varphi)+A^\ast(t)P^\ast(\varphi)+H[A,A^\ast](\varphi),
\label{eq:decomposition}
\end{equation}
with $A(t)=(Q,f_t)_\tau$, $(Q,P^\ast)=0$ and $(Q,H)=0$. We assume that
$H$ is at least of order $(A,A^\ast)^2$. For small $A$ (that is, in the close vicinity of $K_c$),
it can be shown that $R(t)=A^{\ast}(t)+O(|A|^2 A^{\ast}(t))$, so that studying the bifurcation of $A$ is equivalent to that of the order parameter $R$.
Let us define the following Fourier expansions needed for further
analysis:
\begin{eqnarray}
&&f_t=\dfrac{1}{2\pi}\sum_{k=-\infty}^\infty (f_t)_k e^{ik\theta}, 
\\
&&\{\L f_t,\N[f_t]\}=\dfrac{1}{2\pi}\sum_{k=-\infty}^\infty \{\L_k (f_t)_k,\mathscr{N}_k[f_t]\}
e^{ik\theta},\\
&&H\left [A,A^{\ast} \right ]=\dfrac{1}{2\pi}|A|^2 w_0\left [|A|^2 \right ]+\dfrac{1}{2\pi}\sum_{k=1}^\infty \left ( A^k w_k \left [|A|^2 \right ] e^{ik\theta}+(A^\ast)^k w_{-k} \left [|A|^2 \right ] e^{-ik\theta}\right ),
\end{eqnarray}
where the dependence on $A$ of the Fourier coefficients of $H$ is imposed by rotational symmetry \cite{Crawford:1995-1}. 
To proceed with the analysis, we will need to expand the coefficients $w_k$ in powers of $|A|^2$, $w_k=\sum_{j=0}^\infty|A|^{2j} w_{k,j}$. 
To be consistent with the assumption of the unstable manifold being at
least of order $(A,A^\ast)^2$, we need to have $w_{\pm 1,0}=0$.

The Fourier coefficients of the nonlinear operator~(\ref{eq:N}) are
\begin{eqnarray}
\N_k[f_t]&=&\dfrac{i\K}{2 \m}\left (
e^{-i(\alpha+\omega_0\tau)}R_1[f_t](-\tau)\partial_v(f_t)_{k+1}(0)-e^{i(\alpha+\omega_0\tau)}R_{-1}[f_t](-\tau)\partial_v(f_t)_{k-1}(0) \right ).\label{eq:N_k}
\end{eqnarray}
Note that contrary to the case with no inertia, $m=0$, where
$\L_0=\N_0=0$ so that $(f_t)_0=\text{constant}=0$, it is not so in the
present case so that
$(f_t)_0\neq 0$ and $w_0\neq0$.
This difference will have major consequences for the reduction, giving a $1/\lambda$ divergence in the $c_3$ coefficient.

For $\varphi\neq 0$, we find
$w_{0}(\varphi)=h_{0,0}e^{2\lambda_{\rm r}\varphi}+O(|A|^2)$,
$w_{2,0}(\varphi)=h_{2,0}e^{2\lambda\varphi}+O(|A|^2)$, and with the
boundary equation $\varphi=0$:
\begin{eqnarray}
(2\lambda_{\rm r}-\L_0)\cdot h_{0,0} &=& i\dfrac{\K}{2\m}e^{-i(\alpha+\omega_0\tau)}e^{-\lambda^\ast\tau} \partial_v \psi +\cc , \label{eq:del:h00}\\
(2\lambda -\L_2)\cdot h_{2,0} &=& -i\dfrac{\K}{2\m}e^{i(\alpha+\omega_0\tau)}e^{-\lambda\tau}\partial_v \psi \label{eq:del:h20},
\end{eqnarray}
where we used the decomposition, Eq.~\eqref{eq:decomposition}, and
the orthogonal projection with respect to the eigenvectors $\eqref{eq:del:VKdec-0} - (Q,\eqref{eq:del:VKdec-0})P - (Q^\ast,\eqref{eq:del:VKdec-0})P^\ast$ on the Fourier modes $k=0$ and $2$ while only keeping the quadratic orders $O((A,A^\ast)^2)$.
Solving these equation will give us $h_{0,0}$ and $h_{2,0}$ needed in the following.
Projection of the dynamics along the unstable mode using
$(Q,~(\ref{eq:del:VKdec-0}))_\tau$ yields the equation for the amplitude $A(t)$ to be
\begin{eqnarray}
\dot{A} &=& \lambda A + c_3 A |A|^2 +\O{A |A|^4},
\label{eq:del:unstable_mani}
\\ c_3 &=&  \dfrac{ \K}{2i\m} \left
(e^{i(\alpha+\omega_0\tau)}e^{-\lambda\tau} \int \d \omega\,\widetilde{\psi}^\ast \partial_v h_{0,0} - 
e^{-i(\alpha+\omega_0\tau)}e^{-\lambda^\ast\tau}  \int \d \omega \,\widetilde{\psi}^\ast \partial_v h_{2,0}\right ),
\end{eqnarray}
where we used Eq.~(\ref{eq:N_k}) for $k=1$ keeping only the leading order.
To determine the nature of the bifurcation, we must compute explicitly
the coefficient $c_3$. To do that, we must first compute the Fourier component of the unstable manifold.

\subsubsection{Computation of $h_{0,0}$}

We start with Eq.~(\ref{eq:del:h00}). We have $h_{0,0}=h+\cc$, where $h$ is the solution of
\begin{equation}
(2\lambda_{\rm r}-\L_0)\cdot h= i\dfrac{2\pi \K}{2\m}e^{-i(\alpha+\omega_0\tau)}e^{-\lambda^\ast\tau}  \partial_v \psi. 
\label{eq:del:h00bis}
\end{equation}
Equation~(\ref{eq:del:h00bis}) reads
\begin{equation}
    \begin{split}
        &2\lambda_{\rm r} h -\frac{1}{\m}\partial_v[(v-\omega) h]
        =\frac{\K^2}{4i\m^2}e^{-2\lambda_{\rm r}\tau}\left(-\frac{g(\omega)\delta'(v-\omega)}{(\lambda+i\omega)(\lambda+i\omega+1/\m)} +i \frac{g(\omega)\delta''(v-\omega)}{(\lambda+i\omega+1/\m)}\right).
    \end{split}
\end{equation}
We introduce the ansatz
\begin{equation}
    h = W_0(\omega)\delta(v-\omega) + W_1(\omega)\delta'(v-\omega)+
    W_2(\omega)\delta''(v-\omega), 
\end{equation}
to get
\begin{eqnarray}
    W_0(\omega) &=& 0,  \label{eq:del:W0}\\
    W_1(\omega) &=& i\frac{ (\K/2\m)^2
    e^{-2\lambda_{\rm r}\tau}g(\omega)}{(2\lambda_{\rm r}+1
    /\m)(\lambda+i\omega)(\lambda+1/\m+i\omega)}, \label{eq:del:W1}\\
    W_2(\omega) &=& \frac{(\K/2\m)^2 e^{-2\lambda_{\rm r}\tau}
    g(\omega)}{2(\lambda_{\rm r}+1/\m)(\lambda+1/\m+i\omega)}.\label{eq:del:W2}
\end{eqnarray}

\subsubsection{Computation of $h_{2,0}$}

A similar computation starting from Eq.~(\ref{eq:del:h20}) yields $h_{2,0}$.
We have to solve
\begin{equation}
(2\lambda -\L_2)\cdot h_{2,0} = -i\dfrac{ \K}{2\m}e^{i(\alpha+\omega_0\tau)}e^{-\lambda\tau}\partial_v \psi.
\end{equation}
Using the ansatz
\begin{equation}
h_{2,0} = X_0\delta(v-\omega) +X_1\delta'(v-\omega) +X_2\delta''(v-\omega), 
\end{equation}
we obtain
\begin{eqnarray}
X_0(\omega) &=& \frac{iX_1(\omega)}{(\lambda+i\omega)}, \\
X_1(\omega) &=& \frac{- i(
\K e^{i(\alpha+\omega_0\tau)}e^{-\lambda\tau}/2\m)U_0(\omega)}{(2\lambda+2i\omega
+1/\m)}+\frac{4iX_2(\omega)}{(2\lambda+2i\omega+1/\m)}, \\
X_2(\omega) &=& \frac{-i( \K e^{i(\alpha+\omega_0\tau)}e^{-\lambda\tau}/2\m) U_1(\omega)}{2(\lambda+i\omega+1/\m)}.
\end{eqnarray}

\subsubsection{Putting everything together}

One can ascertain that the only diverging term will come from
 $\int \widetilde{\psi}^{(2)\ast}W_1^\ast \,\d \omega$; thus, the leading term is
\begin{equation}
\begin{split}
&\int\d\omega \,\widetilde{\psi}^{(2)\ast} W_1^\ast   \sim  i
\frac{\K^2}{2\m^2}\frac{e^{-2\lambda_{\rm
r}\tau}}{(1/\m)^4}\frac{1}{\Lambda'(i\lambda_{\rm
i})}\frac{\pi}{2}\frac{g(-\lambda_{\rm i})}{\lambda_{\rm r}}.
\end{split}
\label{eq:W1}
\end{equation}
These types of singularities are called ``pinching singularities;''
they arise when two poles approach the real axis, each on one side in an
integral. Indeed, with Eq.~(\ref{eq:del:tildevarpsi}) and the notation
$\widetilde{\psi}(\omega,\omega)=\widetilde{\psi}^{(n)}(\omega)$, we find
that
\begin{equation}
    (\widetilde{\psi}^{(n)})^\ast(\omega) = \dfrac{(-i)^n n!}{\Lambda'(\lambda)}\dfrac{1}{\prod_{l=0}^n (\lambda+i\omega+l/m)}.
\end{equation}
The $(\lambda + i\omega)^{-1}$ factor paired with the $(\lambda^\ast -i\omega)^{-1}$ term appearing only in $W_1^\ast$ gives a ``pinching singularity'' resulting in the $1/\lambda_{\rm r}$ divergence.
We conclude that the leading behavior of $c_3$ for $\m>0$ is given by
\begin{equation}
c_3 \sim \frac{\pi \m\K^3}{8}\frac{e^{i(\alpha+(\omega_0-\lambda_{\rm
i})\tau)}}{\Lambda'(i\lambda_{\rm i})}\frac{g(-\lambda_{\rm
i})}{\lambda_{\rm r}}. 
\label{eq:del:c3eqbis1}
\end{equation}
In particular, the sign of $s(\tau) \equiv \Re\left(\frac{K}{2m}\frac{e^{i(\alpha+(\omega_0-\lambda_{\rm
i})\tau)}}{\Lambda'(i\lambda_{\rm i})}\right)$
determines the type (sub- or super-critical) of the bifurcation.

Summarizing the analysis of this subsection, we find the following reduced equation for the order
parameter: 
\begin{eqnarray}
\dot{A}&=&\lambda A + c_3(\lambda) |A|^2 A+O(|A|^4 A),\label{eq:A_red}\\
c_3(\lambda) &\sim & \pi m
\dfrac{K^3}{8\gamma^4}\frac{e^{i(\alpha+(\omega_0-\lambda_{\rm
i})\tau)}}{\Lambda'(i\lambda_{\rm i})}\frac{G(\omega_0-\lambda_{\rm
i})}{\lambda_{\rm r}}, \quad \lambda_{\rm r}\to 0^+,
\label{eq:del:c3eqbis}
\end{eqnarray}
where the unstable eigenvalue $\lambda$ is decomposed into its real and
imaginary parts: $\lambda=\lambda_{\rm r}+ i\lambda_{\rm i}$.
A few remarks are in order:
a) The coefficient $c_3$ diverges as $\lambda_{\rm r}\to 0$, which is the
regime where the reduction is valid. This singular behavior is typical
of this type of systems~
\cite{Crawford:1995,Crawford:1995-1,Barre:2016}, and stems from the
existence of the continuous eigenspectrum that cannot be described by
the finite dimensional equation~(\ref{eq:A_red}).
b) However, we still expect the
behavior of $c_3$ to determine the type of bifurcation.
For $\Re(c_3)>0$, we expect a subcritical (discontinuous) bifurcation, while
for $\Re(c_3)<0$, we expect a supercritical bifurcation. 
In the latter case, the scaling of the stationary amplitude is $A_{\rm{st}}\propto
\lambda_{\rm r}$, which differs from the usual Kuramoto model where it
goes as $\sqrt{\lambda_{\rm r}}$.
c) In the case with no inertia, that is, with $m=0$, we expect the
coefficient $c_3$ to be quantitatively relevant in giving the exact
amplitude $A_{\rm{st}}$ of the stationary branch close to the bifurcation;
here, because of the singularity, only the sign and scaling of $c_3(\lambda)$ can be used heuristically to get qualitative information.  
Heuristically, the unstable manifold procedure will describe the
linear growth of the instability until the nonlinear effects, governed
by $c_3$, kick in, and then the simple one-dimensional reduced model,
Eqs.~(\ref{eq:A_red},\ref{eq:del:c3eqbis}), cannot capture the full saturation dynamic.
However, even if the diverging term in Eq.~\eqref{eq:W1} is always the dominating contribution in $c_3(\lambda)$ for $m\neq 0$, one can intuitively guess that away from the bifurcation point $K \gtrsim K_c$, this term proportional to $m$ will become `very quickly' small compared to other terms contributing to $c_3(\lambda)$ when $m$ is small. Hence, one can expect that other very different bifurcations take place closely after the bifurcation. In practice, this is what we observe for smaller $m$ in numerical simulations, see Fig.~\ref{fig:num_results_m0p1}.
\subsection{Application to a Lorentzian distribution}

\begin{figure}[!htbp]
    \begin{center}
        \includegraphics[width=0.7\textwidth]{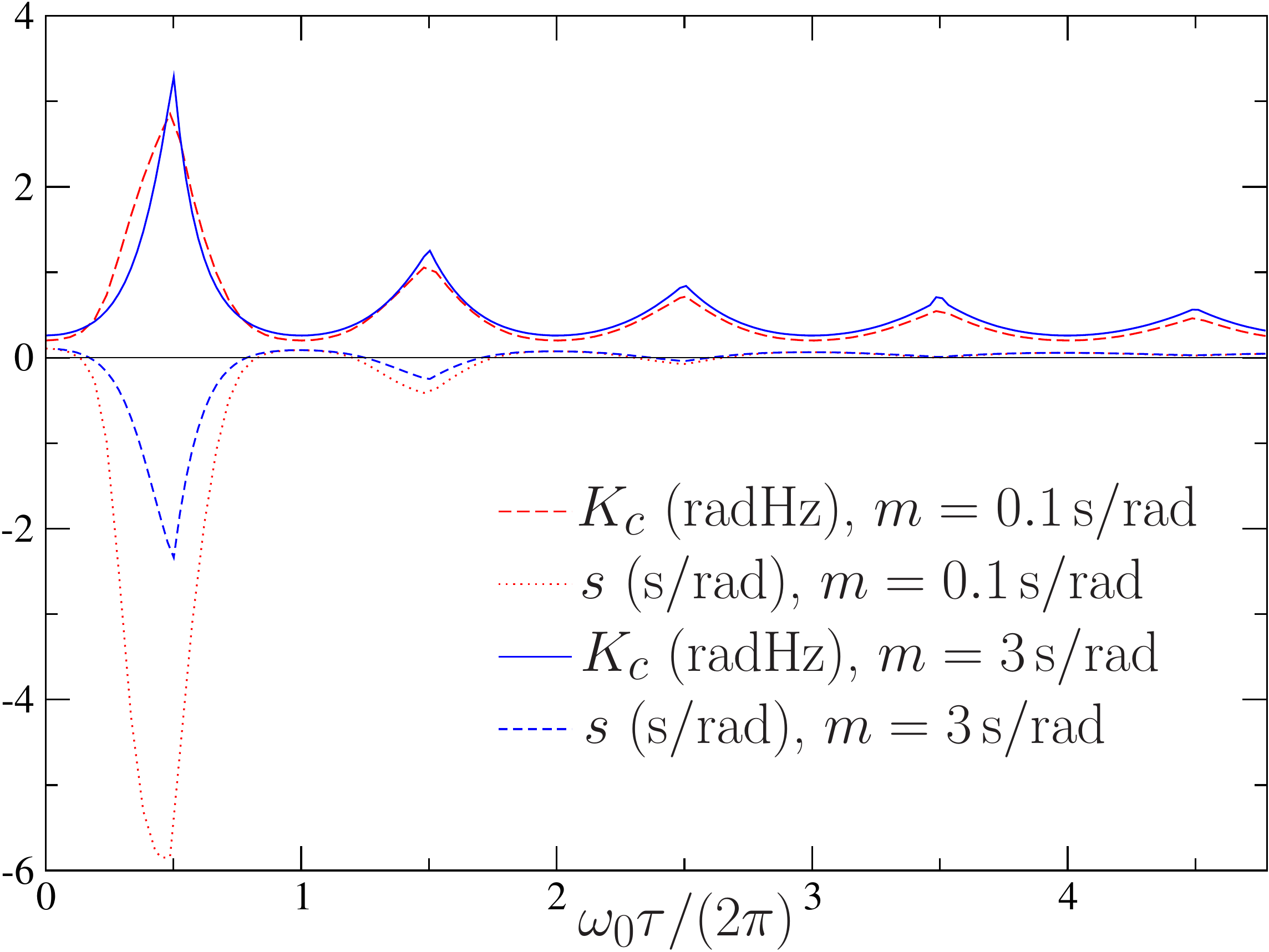}
        \caption{Stability region of the incoherent state for Lorentzian $G_L (\omega) = \sigma /[\pi ((\omega - \omega _0 )^2 + \sigma^2 )]$ with $\sigma = \SI{0.1}{rad Hz}$, $\omega_0 = \SI{3}{rad Hz}$, $\gamma = 1$ and $\alpha = 0$. We show here as a function of $\tau$ the quantities $K_c(\tau )$ and $s(\tau )$ for $m = \SI{0.1}{s/rad}$ and $m = \SI{3}{s/rad}$. The sign of $s(\tau )$, satisfying $\operatorname{sign} (\Re(c_3 )(\tau )) = \operatorname{sign} (s(\tau ))$, determines the super- or subcritical nature of the bifurcation.}
        \label{fig:del:strog_m}
    \end{center}
\end{figure}

To assess the effects of inertia in a delay system, we consider a Lorentzian distribution of the  natural frequencies: $G_L(\omega)=\sigma/[\pi ((\omega-\omega_0)^2+\sigma^2)]$.
The dispersion relation~(\ref{eq:del:Lambda}) at criticality gives $K=K_c$ and $\lambda=0^++i\lambda_{{\rm i,c}}$:
\begin{eqnarray}
    &&\dfrac{K_c}{2}=\left (\gamma\sigma+m\sigma^2-m\lambda_{{\rm i,c}}^2\right
    )\sec((\lambda_{{\rm i,c}-\omega_0})\tau), \label{eq:sys_K}\\
    &&\dfrac{\lambda_{{\rm i,c}}(\gamma+2\sigma
    m)}{\gamma\sigma+m\sigma^2-m\lambda_{{\rm
    i,c}}^2}=-\tan((\lambda_{{\rm i,c}-\omega_0})\tau),
    \label{eq:sys_L}
\end{eqnarray}
where for simplicity, we chose $\alpha=0$ for this application.
Solving this system gives us $K_c(\tau)$ and $\lambda_{{\rm i,c}}(\tau)$.
Then the sign of the cubic coefficient $\Re(c_3)(\tau)$ is given by
\begin{equation}
s(\tau)=\Re\left (\dfrac{(\sigma +i \lambda_{{\rm i,c}} )^2 (\gamma+m (\sigma
+i \lambda_{{\rm i,c}} ))^2}{m (\gamma+(\sigma +i \lambda_{{\rm i,c}} ) (\gamma\tau
+m (i \lambda_{{\rm i,c}}  \tau +\sigma  \tau +2)))} \right ),
\end{equation}
where we used Eq.~(\ref{eq:del:c3eqbis}) with $G=G_L$.
We plot in Fig.~\ref{fig:del:strog_m} the quantities $K_c$ and $s$ as a
function of the delay for two different inertia values $m=\SI{0.1}{s/rad}$ and $m=\SI{3}{s/rad}$
for a Lorentzian distribution. 
For $\tau\to 0$, we recover the no-delay results showing a positive
$\Re(c_3)$ and hence a subcritical
bifurcation~\cite{Gupta-Campa-Ruffo:2014-2}. 
Moreover, as in the case with no inertia~\cite{Metivier:2019}, $m=0$, the delay induces ``oscillations" in the sign of $\Re( c_3)$. 
We observe that different nonzero values of $m$ do not change much the behavior of the bifurcation.

\section{Numerical results}
\label{sec:numerics}

\subsection{Method}

In the preceding section, we provided an analytic characterization of the stability properties of the incoherent state in the Kuramoto model with delayed coupling and inertia in the limit $N\to\infty$.
Here we present results from numerical integration of the dynamics~(\ref{eq:Kuramoto-inertia-delay-1-rt}) with Lorentzian-distributed natural frequencies with location parameter $\omega_0=\SI{3}{radHz}$ and scale parameter $\sigma=\SI{0.1}{radHz}$.
For all-to-all coupling, Eq.~(\ref{eq:Kuramoto-inertia-delay-1-rt}) can be rewritten in terms of the Kuramoto order parameter:
Using $R_x(t-\tau)=1/N\sum_j\,\cos(\theta_j(t-\tau))$ and $R_y(t-\tau)=1/N\sum_j\,\sin(\theta_j(t-\tau))$, we rewrite the equations of motion as
\begin{eqnarray}
\dot{\theta}_i(t)&=&v_i(t), \nonumber \\
\label{eq:Kuramoto-inertia-meanfield_R} \\
m\dot{v}_{i}(t)+\gamma
v_{i}(t)&=&\gamma\omega_{i}+K\,\left[R_y(t-\tau)\cos(\theta_i(t)) - R_x(t-\tau)\sin(\theta_i(t))\right]\nonumber .
\end{eqnarray}
%
We set $\alpha = 0$ for the numerical experiments. 
Hence, for $\gamma = 1$, we have the set of equations
\begin{eqnarray}
\dot{\theta}_i(t)&=&\omega_i + K\,x_i^{\rm C}(t), \\
\dot{x}_i^{\rm C}(t)&=&\frac{1}{m}\left( x_i^{\rm PD}(t) - x_i^{\rm C}(t) \right), \\
x_i^{\rm PD}(t) &=& R_y(t-\tau)\cos(\theta_i(t)) - R_x(t-\tau)\sin(\theta_i(t)),
\end{eqnarray}
which we integrate numerically using an Euler iteration-method, given the initial phases $\theta_i(0)$ independently and identically distributed in $\left[0, 2\pi\right)$, and the  initial states of the filters $x_i^{\rm C}(0)$.
$R_x(t_{\rm hist})$ and $R_y(t_{\rm hist})$ with $t_{\rm hist}\;\in\;\left[-\tau,\,0\right]$ given by the history of the network of oscillators, which we obtain by evolving each oscillator independently according to its own natural frequency, i.e., as if they were uncoupled.
The code is written in python and compiled with cython for fast execution and can be found on the Gitlab  repository here~\cite{source_code_GH}.

\begin{figure}[!htbp]
\begin{flushleft}
\includegraphics[width=0.45\textwidth]{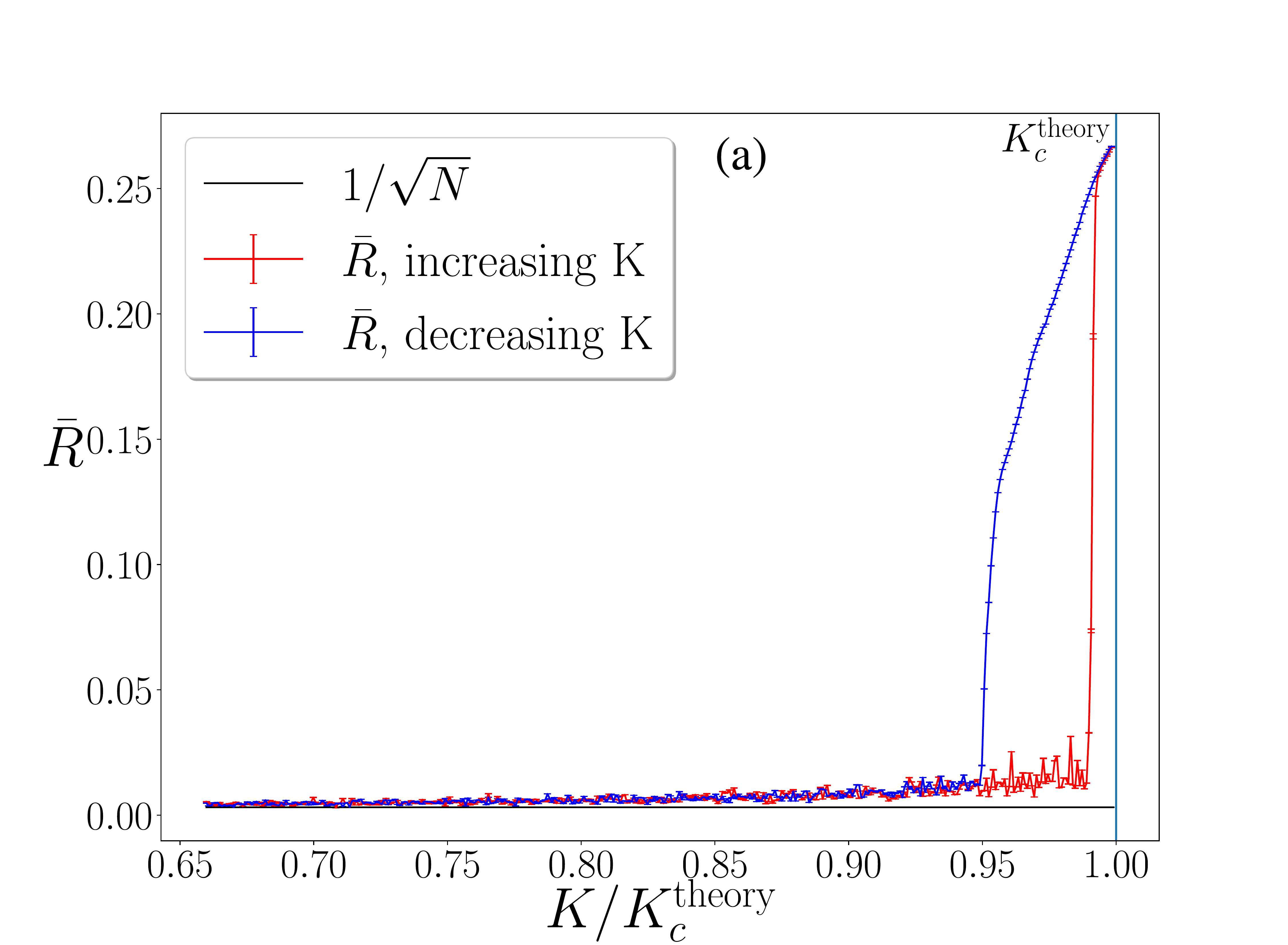}
\includegraphics[width=0.45\textwidth]{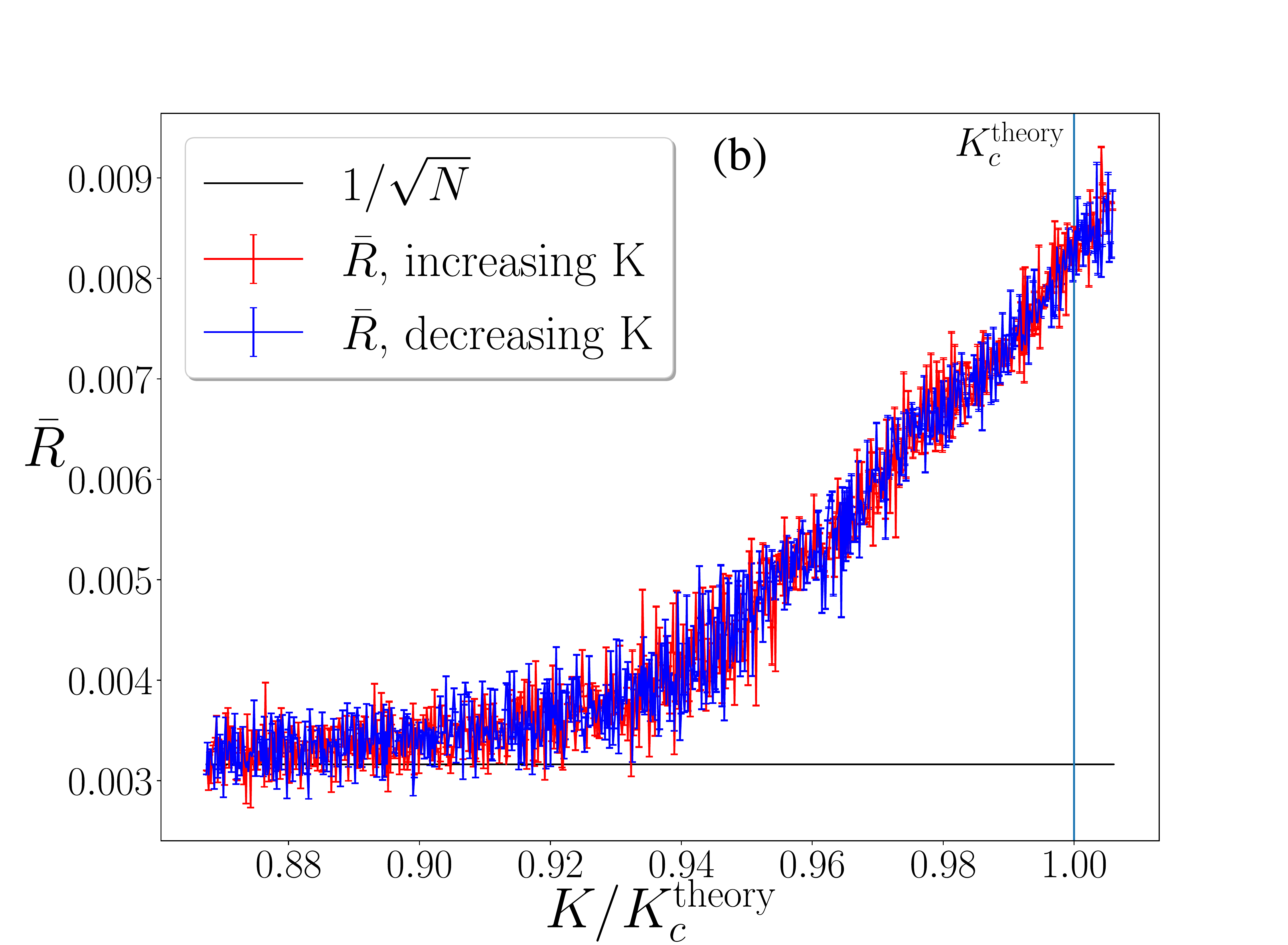} \\
\includegraphics[width=0.45\textwidth]{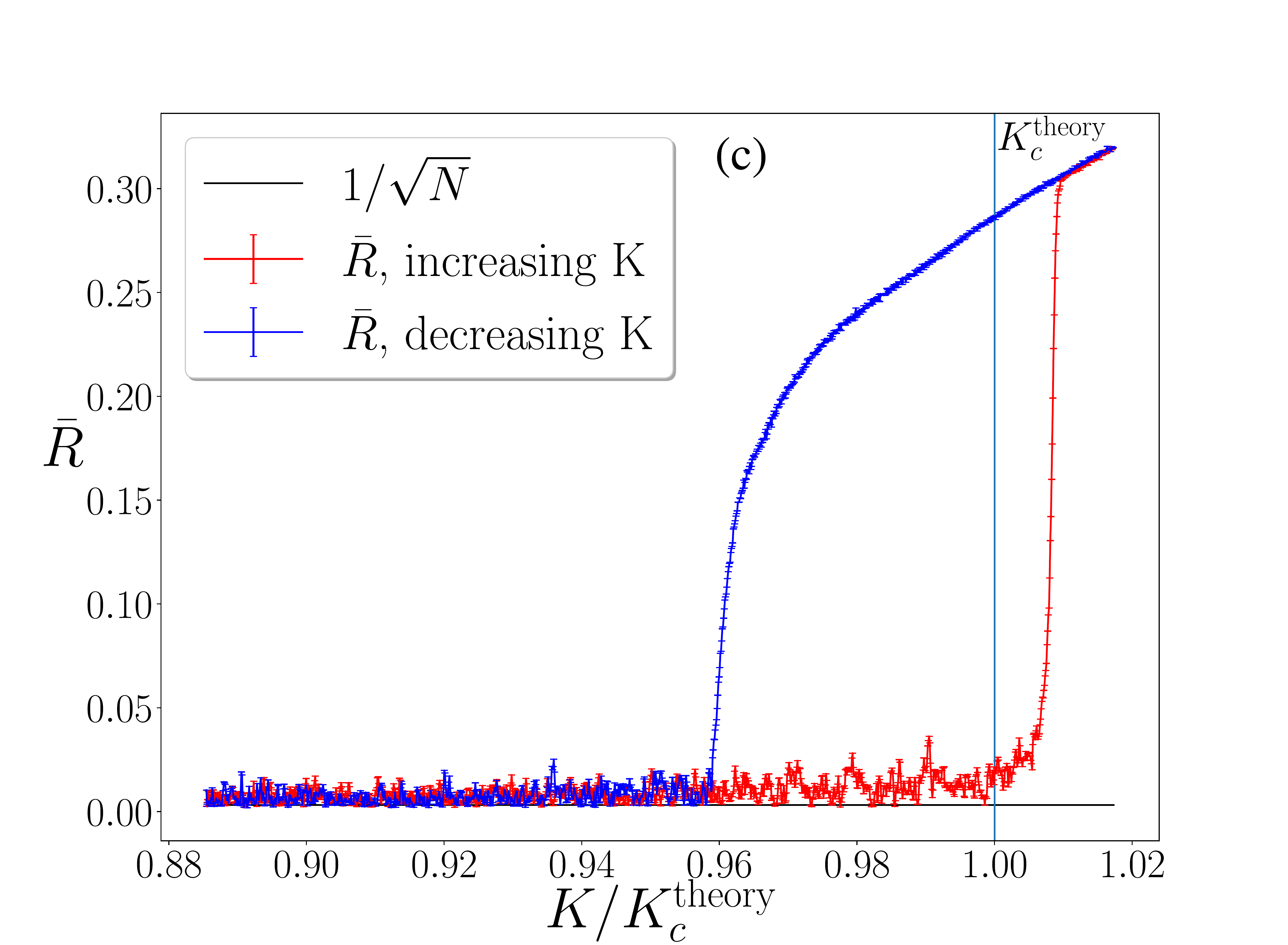}
\includegraphics[width=0.45\textwidth]{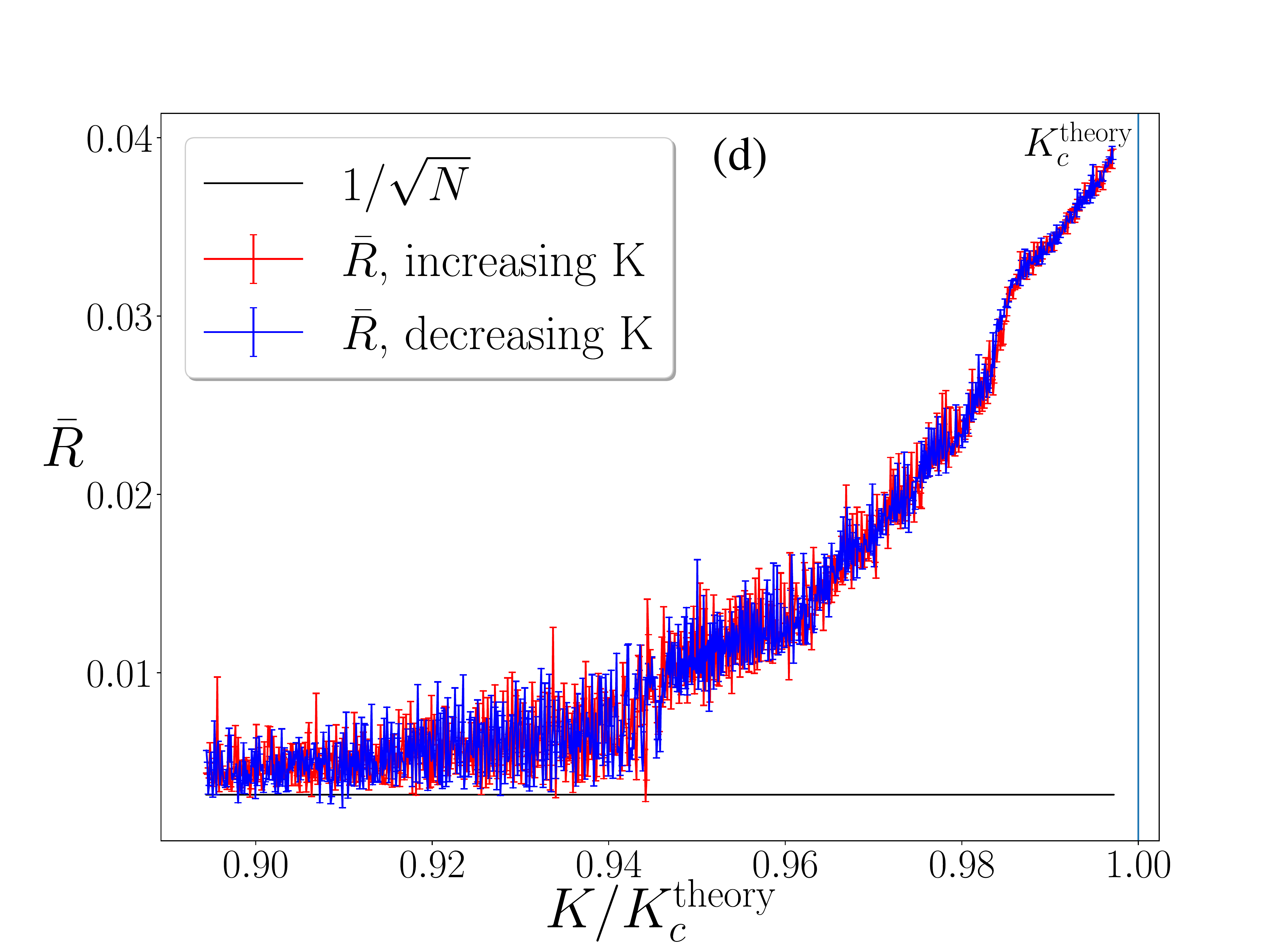} \\
\includegraphics[width=0.45\textwidth]{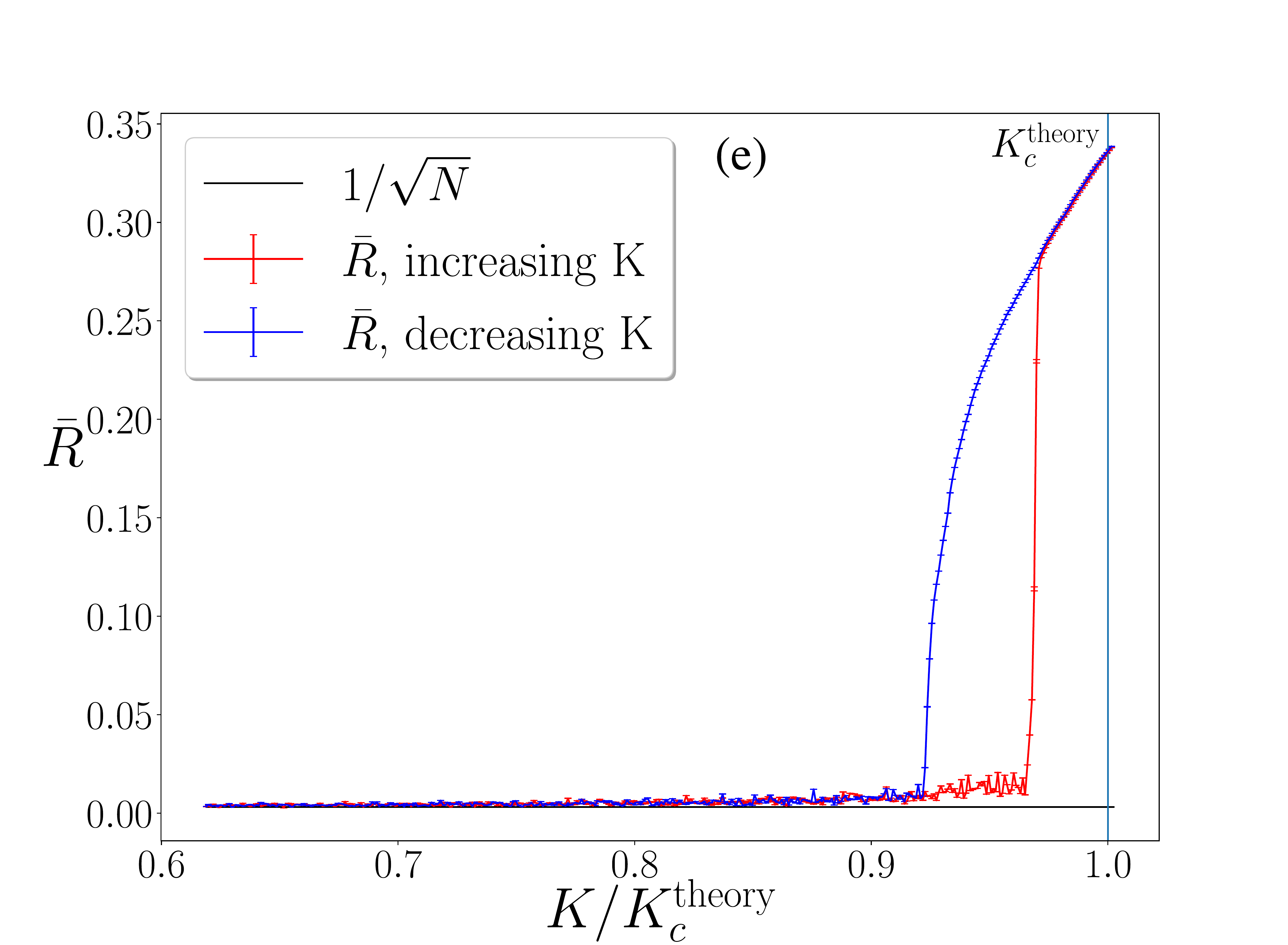}
\includegraphics[width=0.45\textwidth]{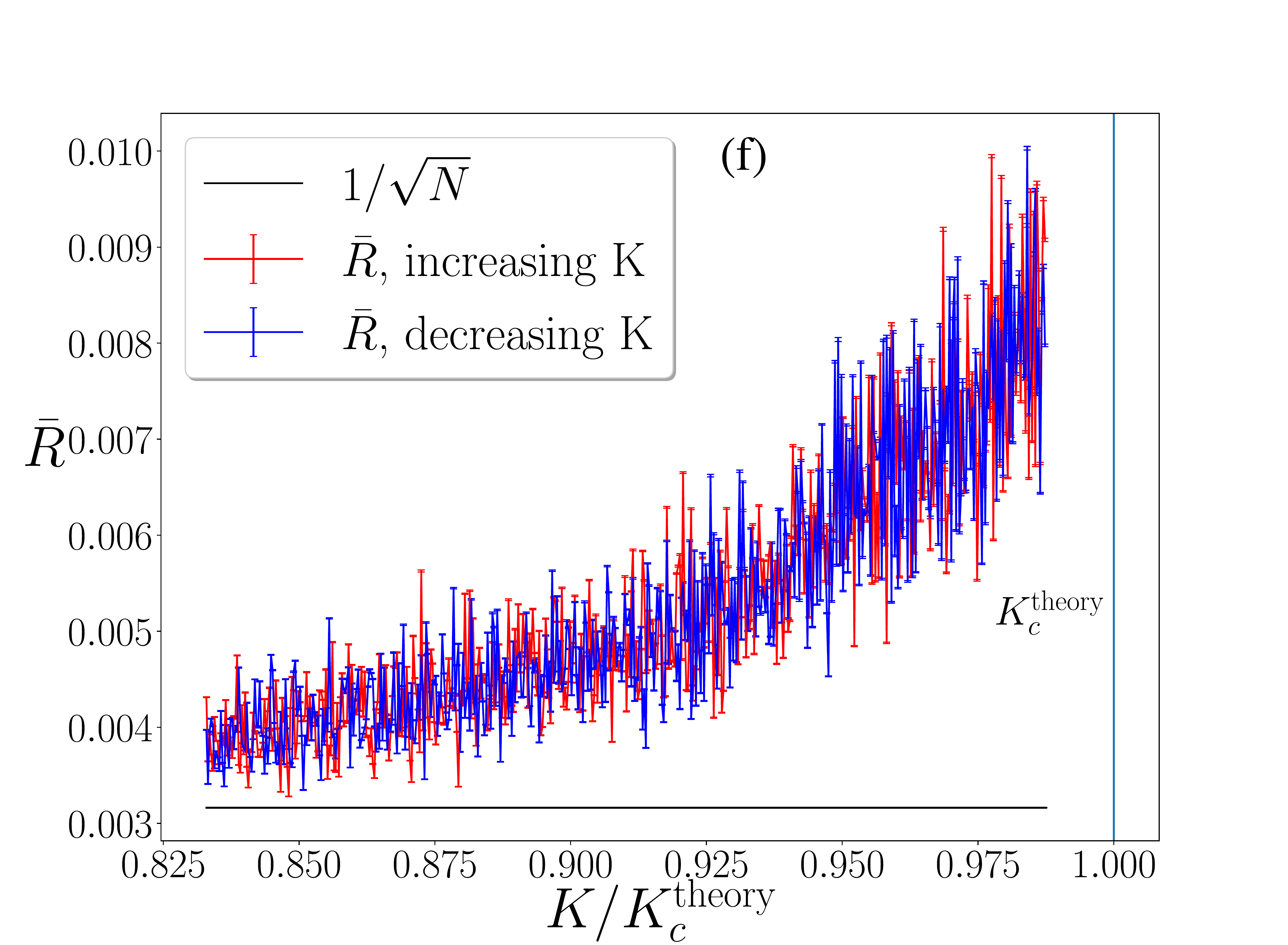}
\caption{Numerical integration results for the average Kuramoto
order-parameter $\bar{R}(t)$ as a function of the coupling constant $K$
in the vicinity of the critical coupling constant $K_c$. Plots show the
cases with inertia $m=\SI{3.0}{s/rad}$ and
$\tau=\{0.2,\,1.0,\,2.0,\,3.0,\,4.2,\,5.25\}\,\si{s}$ in order from (a) to (f). For each
value of $K$, the system of $N=10^5$ oscillators was integrated for
at least $T_{\rm num}=\SI{200}{s}$,
and the $K$-values are separated by $\Delta K=(2.5\times
10^{-4})/(2\pi)\,\si{Hz}$, see text. The horizontal black lines at $\bar{R}=1/\sqrt{N}$ denote the order parameter fluctuations expected for a finite system with $N$ unsynchronized oscillators.
}
\label{fig:num_results_m3}
\end{flushleft}
\end{figure}
Our objective behind performing the numerics is to verify our theoretical results obtained in the limit $N \to \infty$ for the critical coupling constant $K_c$ above which the incoherent state becomes unstable.
Furthermore, we want to confirm the type of bifurcation as predicted by our theoretical results that can be observed as the coupling constant $K$ is tuned across $K_c$.
To this end, we integrate numerically the dynamics of large networks of all-to-all delay-coupled Kuramoto oscillators in the vicinity of the theoretically-predicted $K_c$, see Fig.~\ref{fig:del:strog_m}.

We proceed with our numerical work as follows.
For a given set of parameters $(N, \tau, m, \gamma=1, \alpha=0)$, a set of discrete coupling constants $K(n)=\{ K_{\rm on},\,K_{\rm on}+\Delta K,\,\dots,\,K_{\rm end}-\Delta K,\,K_{\rm end} \};~~\Delta
K>0$ and a set of Lorentzian-distributed natural frequencies $\{\omega_i\}$, each oscillator is evolved independently with its own natural frequency for a time $\tau$ to obtain the dynamical history for $N$ oscillators in the interval $[-\tau, 0]$.
In the next step, we turn on the coupling between the oscillators at an initial coupling constant $K_{\rm on}$ that is close to but smaller than the critical coupling constant predicted by our theoretical results. 
Subsequently, the delay-coupled system of all-to-all coupled oscillators is
evolved with the coupling constant kept fixed at $K_{\rm on}$ for time
$T_{\rm num}$ that is long compared to the mean period of the
independent oscillators, in order to ensure that the system settles into
a stationary state at the fixed value of the coupling.
Then, using the phases of the final interval $[T_{\rm num}-\tau, T_{\rm num}]$ as the history, we evolve the system of coupled oscillators for the next larger value in $K(n)$ for time $T_{\rm num}$, and so on, until the final value $K_{\rm end}$ is reached.
In the final part of this procedure, we follow the exact reverse protocol, namely, repeating the above steps while decreasing the value of the coupling from $K_{\rm end}$ until the initial value $K_{\rm on}<K_c$ is reached.
In numerics, we track the value of the Kuramoto order parameter in time,
and save for each value of the coupling in the set $K(n)$ the final value of the order parameter obtained at the end of run for time $T_{\rm num}$ as well as its average and variance computed over a time $t_{\rm average}$ equal to $50$ times the time period corresponding to $\omega_0$, i.e., $t_{\rm average}=50\times2\pi/\omega_0$.

\subsection{Results}

\begin{figure}[!htbp]
\begin{flushleft}
\includegraphics[width=0.45\textwidth]{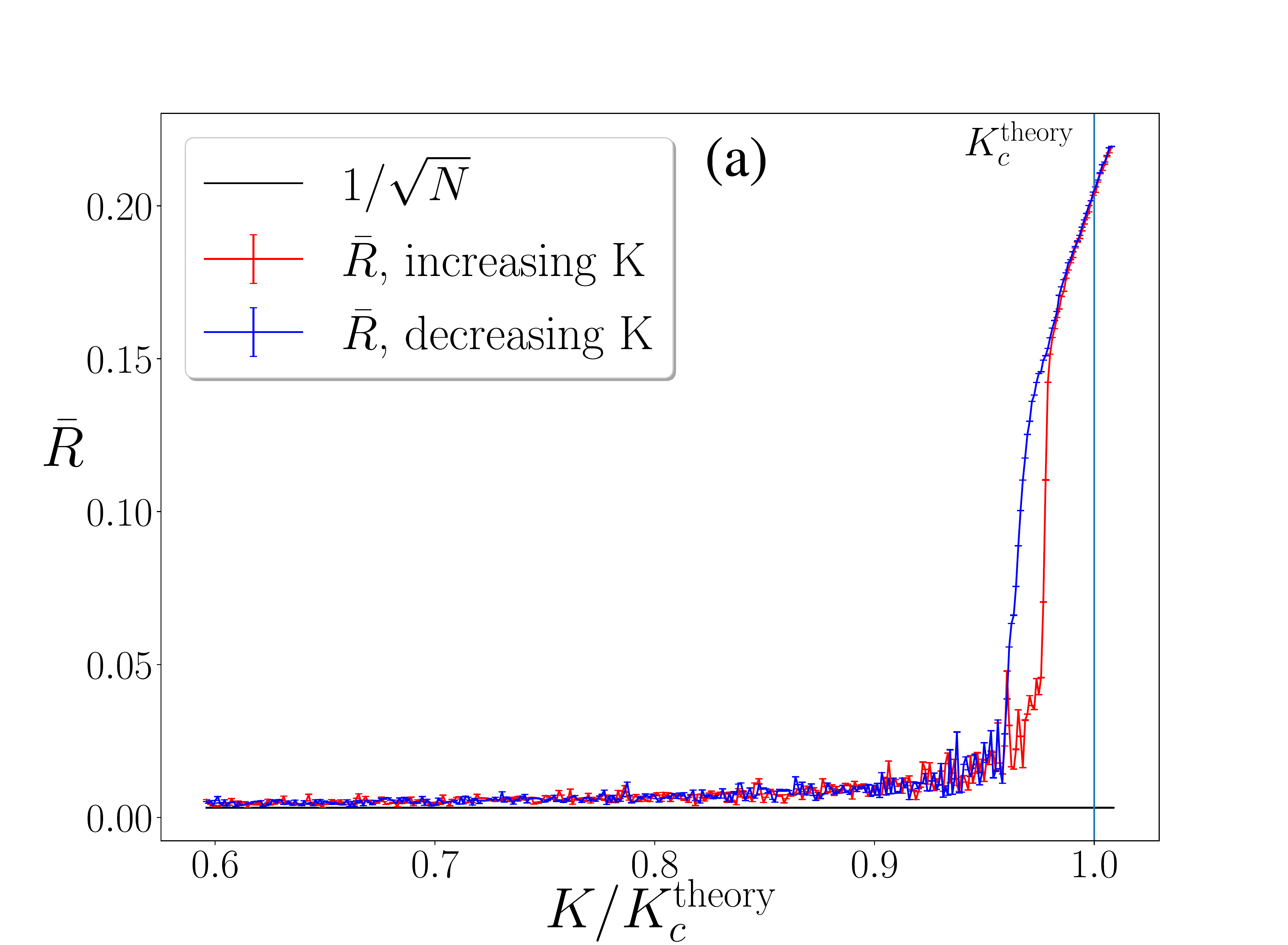}
\includegraphics[width=0.45\textwidth]{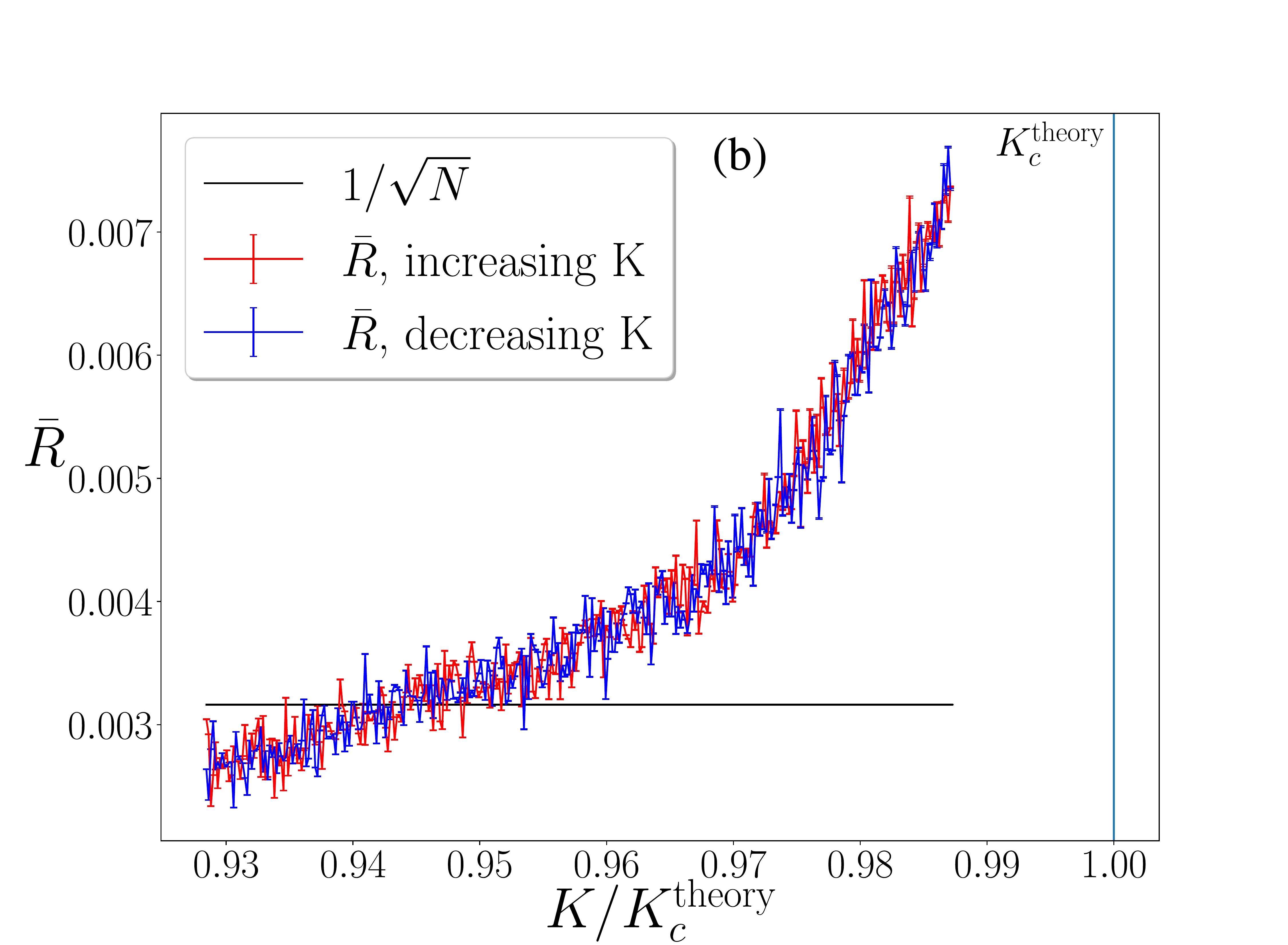} \\
\includegraphics[width=0.45\textwidth]{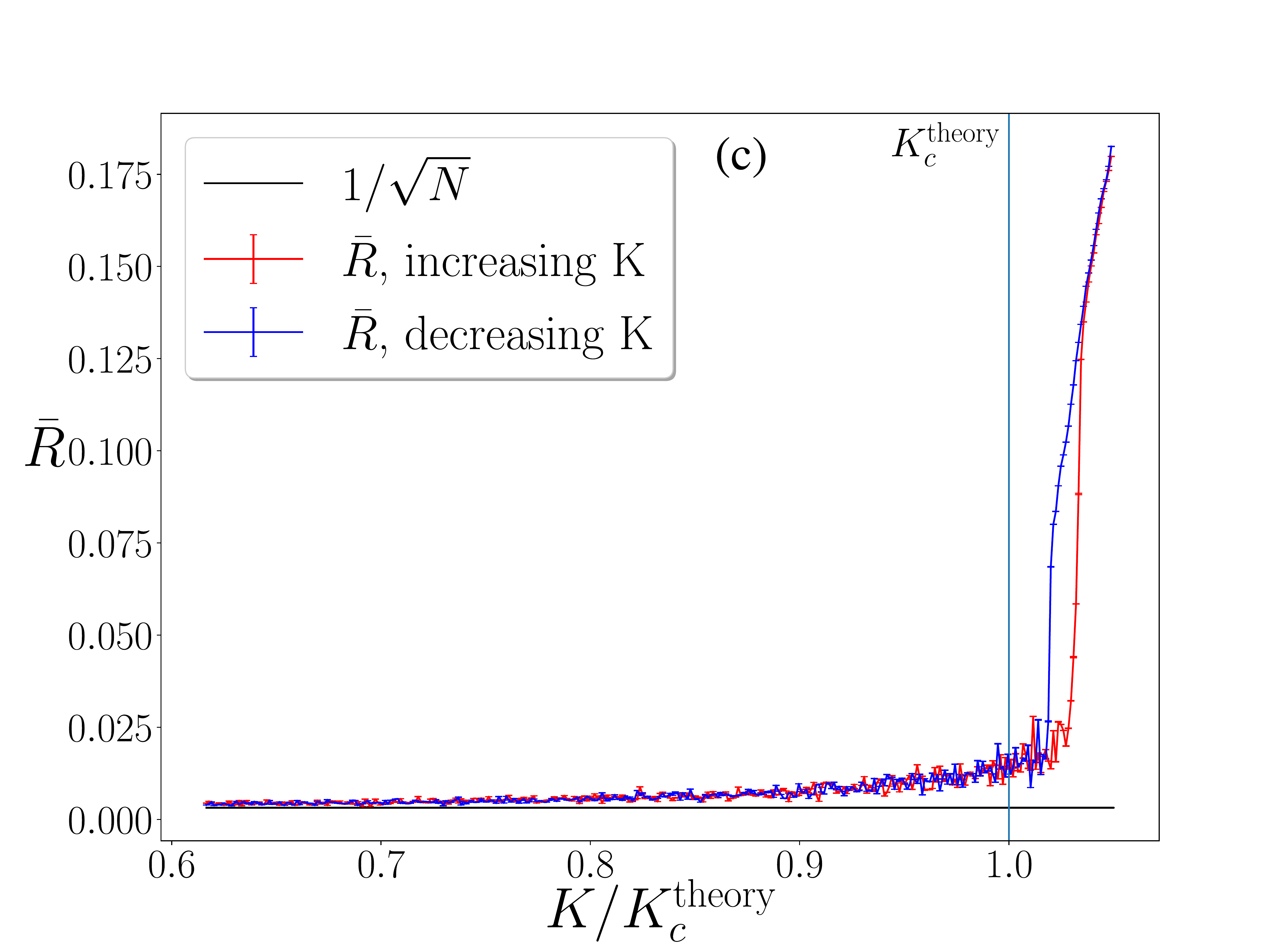}
\includegraphics[width=0.45\textwidth]{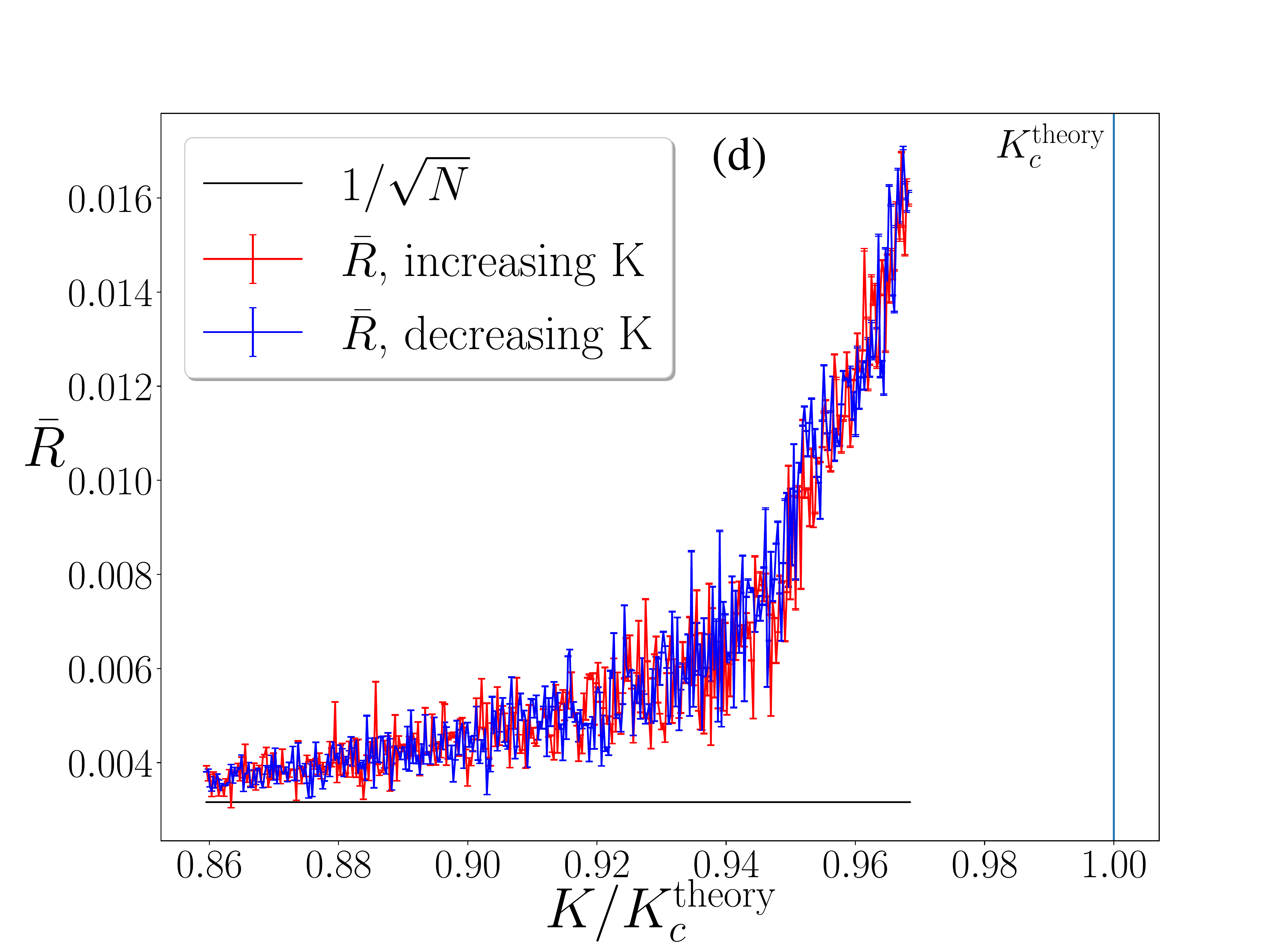} \\
\includegraphics[width=0.45\textwidth]{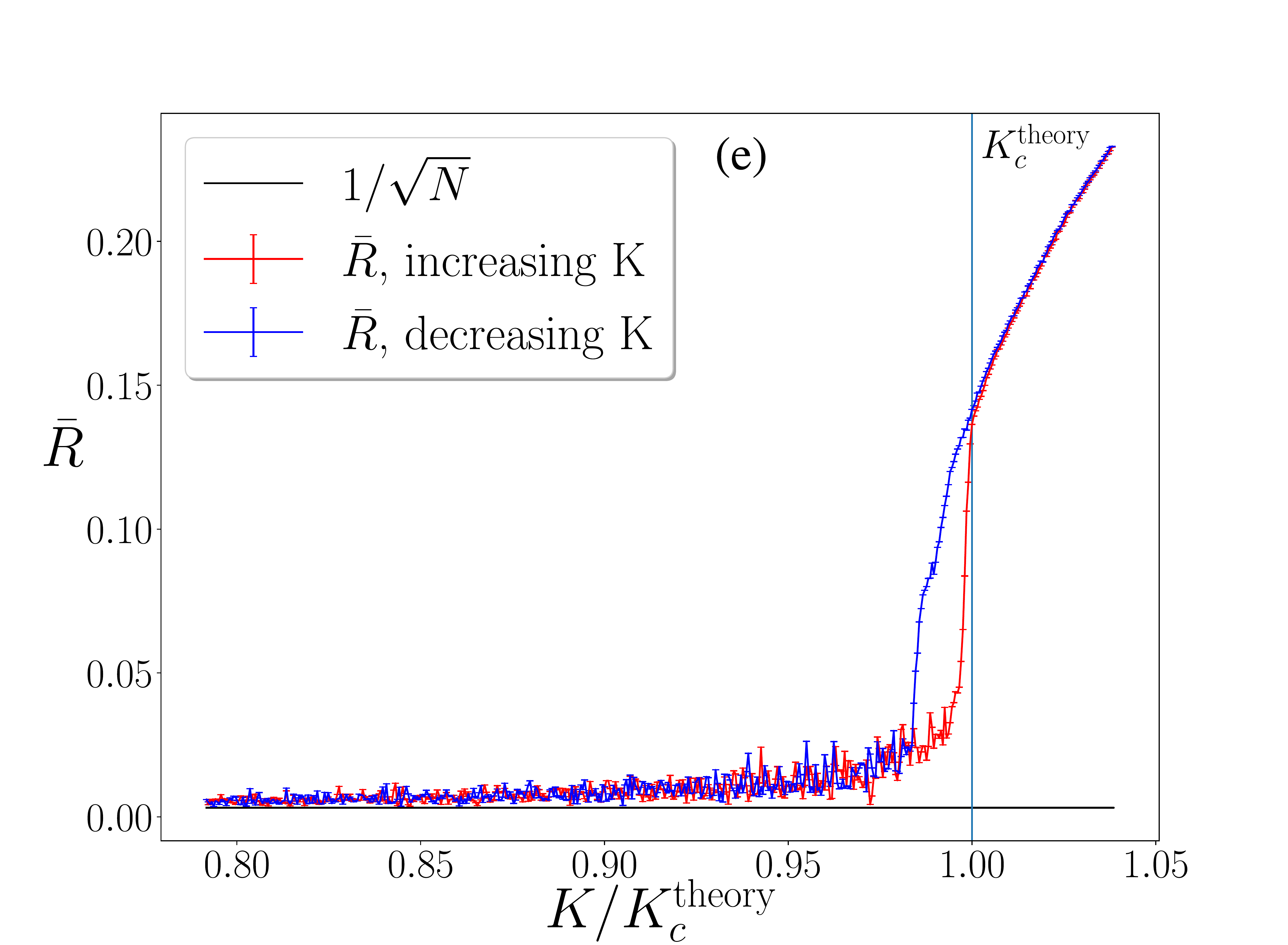}
\includegraphics[width=0.45\textwidth]{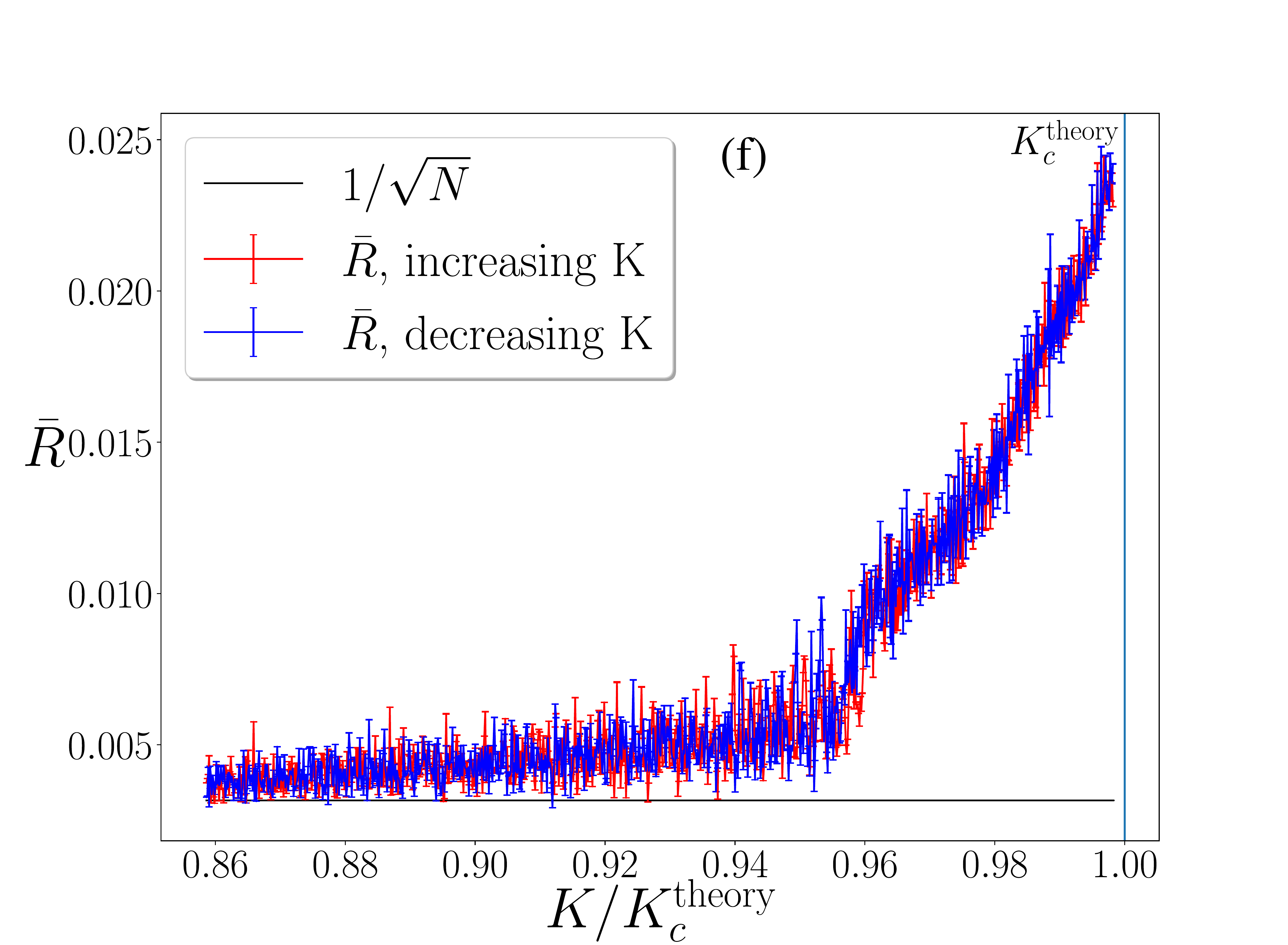}
\caption{Numerical integration results as in Fig.~\ref{fig:num_results_m3} but for $m=\SI{0.1}{s/rad}$.
}
\label{fig:num_results_m0p1}
\end{flushleft}
\end{figure}

We present in Figs.~\ref{fig:num_results_m3} and~\ref{fig:num_results_m0p1} (a)-(f) results for transmission delays $\tau=\{0.2,\,1.0,\,2.0,\,3.0,\,4.2,\,5.25\}\,\si{s}$ and moments of inertia  $m=\{3.0,\,0.1\}\,\si{s/rad}$, obtained for a system of $N=10^5$ all-to-all coupled oscillators.
In both the figures, the left panels (respectively, right panels) show the cases for which the theory predicts $\Re(c_3)>0$ and hence a
subcritical bifurcation and presence of a hysteresis loop (respectively, $\Re(c_3)<0$ and hence a supercritical bifurcation with no hysteresis). 
The plots show the Kuramoto order parameter averaged over a time equal to $50$ times the time period corresponding to the frequency frequency $\omega_0$ and plotted as a function of the coupling constant $K$.

%
In our simulations, we find the bifurcations that were predicted by the theoretical results. We denote by $K_c^{\rm
theory} \equiv K$ the theoretical critical coupling predicted by Eqs.~(\ref{eq:sys_K},\ref{eq:sys_L}) in the Lorentzian case $G(\omega)=G_L(\omega)$.
As the coupling constant $K$ increases, it crosses a critical coupling constant $K_c^{\rm num}$ as found in our finite-size simulation, and we observe a subcritical (discontinuous) transition and hysteresis for $\Re(c_3)>0$. 
$K_c^{\rm num}$ denotes the value of the coupling strength in the numerical calculations at which the incoherent state becomes unstable.
%
%
For $\Re(c_3)<0$, on the other hand, we find a supercritical (continuous) transition with a linearly growing order parameter and no hysteresis as $K$ grows larger than $K_c^{\rm num}$.
We observe that for the case of $m=\SI{0.1}{s/rad}$, the hysteresis seems to be weaker than in the case with $m=\SI{3}{s/rad}$. 
The obtained results are in good agreement with our theoretical predictions for $K_c$ and the type of bifurcation.  

\subsection{Discussion of numerical results}

Numerical validation of our analytical $N \to \infty$-limit results obtained is not trivial and comes with a few difficulties that we will discuss here.
Since we do not know the critical system size $N_{\rm crit}$ of oscillators below which strong finite-size effects will come into play nor the number $N_{\rm thermo}$ above which the behavior coincides with that in the thermodynamic limit, we decided to go for as large a system size as is practicable. 
This however becomes very resource intensive, since for systems with time delays, a memory of the states of all oscillators for a time period $[t-\tau, t]$ has to be stored in order to perform dynamical evolution.
For the mean-field coupling case, we have the advantage that it is sufficient to store only the history of the order parameter variables $R_x(t-\tau)$ and $R_y(t-\tau)$.

\begin{figure}[!t]
\begin{flushleft}
\includegraphics[width=0.45\textwidth]{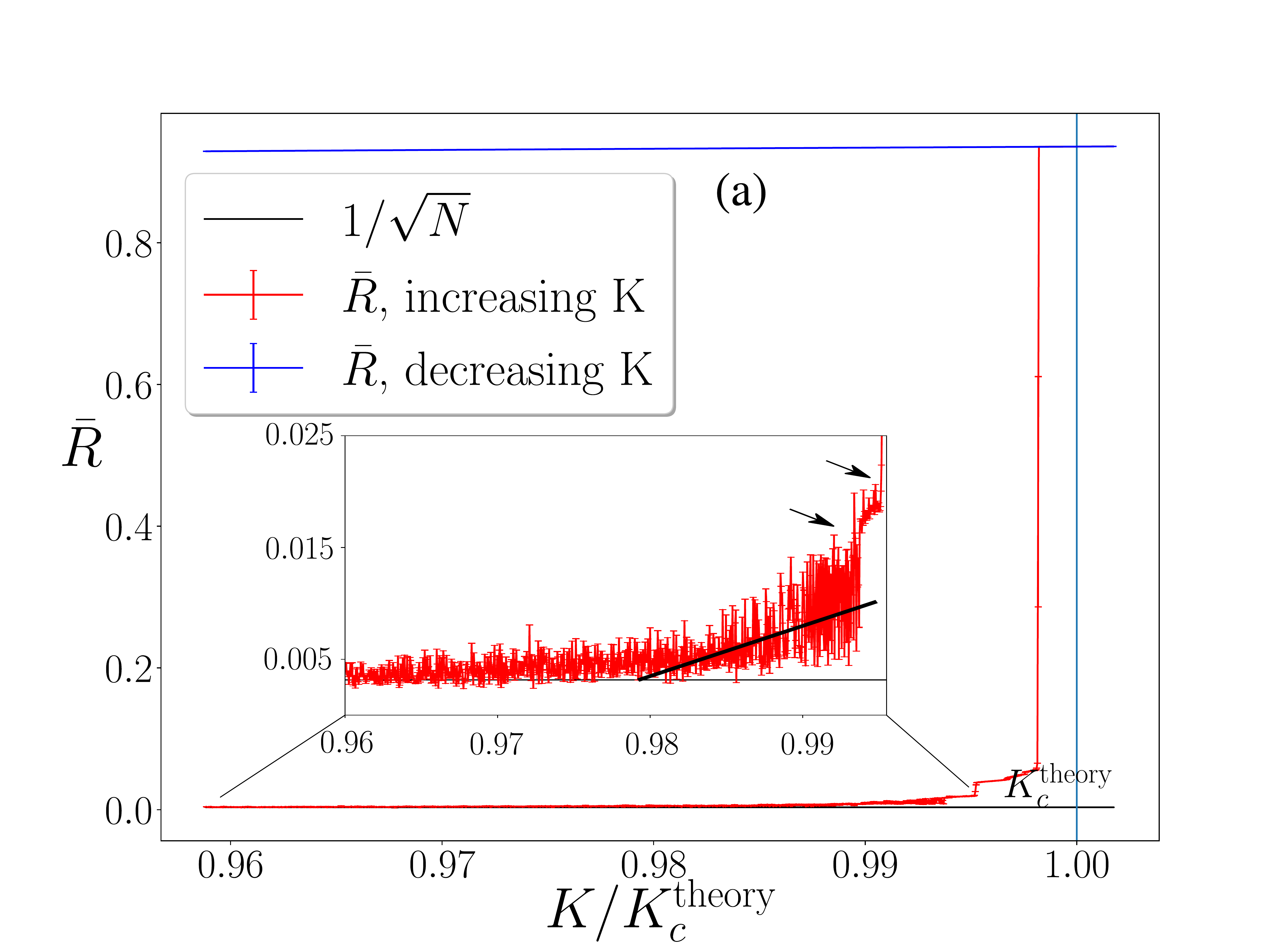} 
\includegraphics[width=0.45\textwidth]{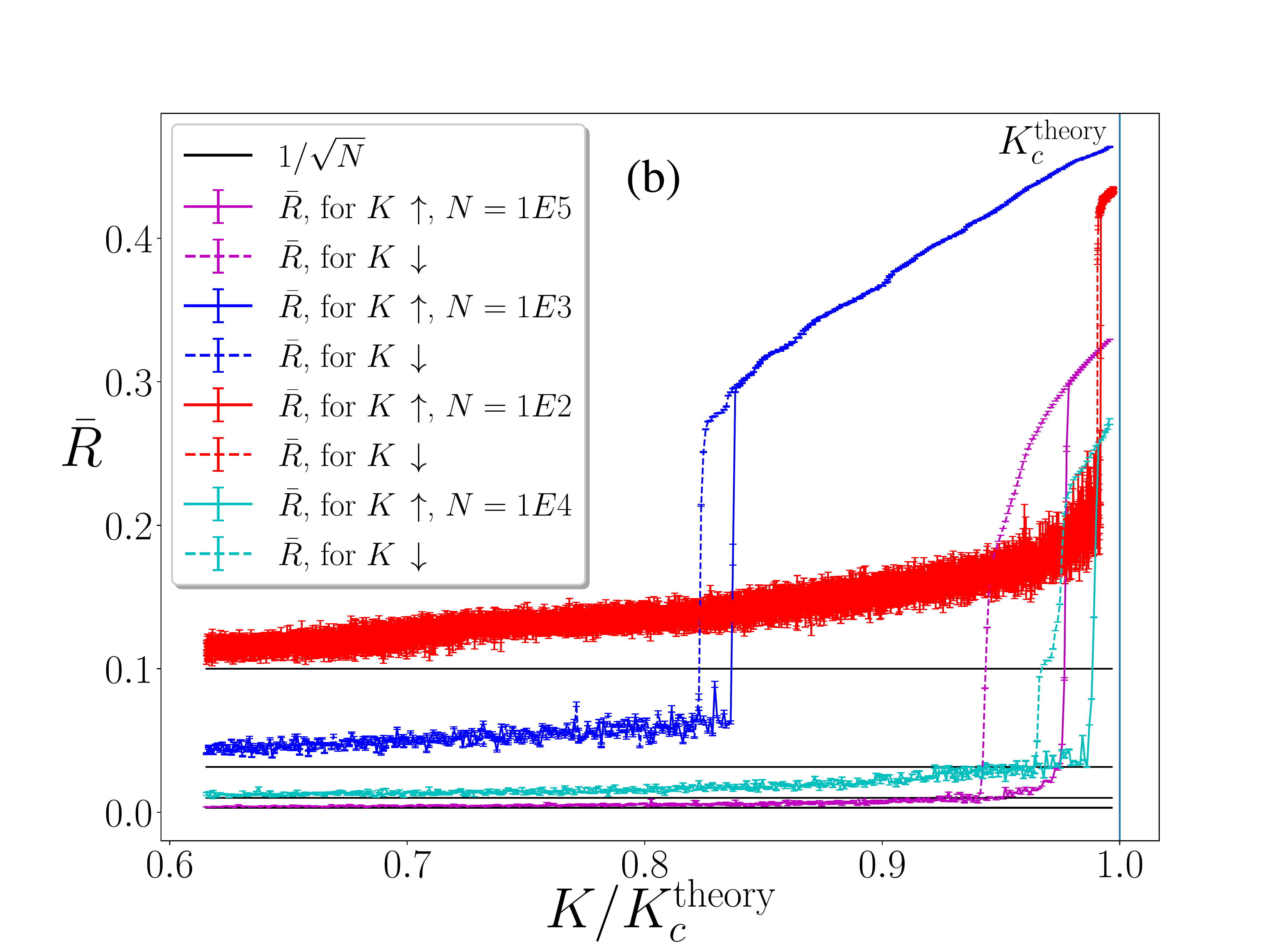}
\caption{Plot (a) addresses the difficulties in showing the supercritical bifurcation arising due to subsequent bifurcations close to the critical coupling strength for the case $m=\SI{0.1}{s/rad}$ and $\tau=\SI{1}{s}$. Plot (b) addresses the issue of finite-size effects in our simulations, and we show results for different numbers $N$ of oscillators for the case $m=\SI{3}{s/rad}$ and $\tau=\SI{4.2}{s}$ The arrows in the legend denote whether $K$ is being increased or decreased.}
\label{fig:num_results_supp1}
\end{flushleft}
\end{figure}

It is known from the literature \cite{Hong:2007} that for the Kuramoto
model in absence of delay and inertia, when the number of oscillators is
finite, we may expect to find $K_c^{\rm num} \leq K_c^{\rm
theory}$, depending on the number $N$ of oscillators considered. 
%
%
This difference is even stronger when considering the inertial model and
subcritical bifurcation, where the convergence $K_c^{\rm theory}-K_c^{\rm num}(N)\sim N^{-0.22}$ \cite{Olmi:2014} is
very slow with $N$ (compared with $K_c^{\rm theory}-K_c^{\rm
num}(N)\sim N^{-0.4}$ without inertia \cite{Hong:2007}). 
Note that the prediction $K_c^{\rm theory}>K_c^{\rm num}$ was obtained for systems without delay, thus observing $K_c^{\rm theory} \approx K_c^{\rm num}$ in Fig.~\ref{fig:num_results_m3} and Fig.~\ref{fig:num_results_m0p1} at $\tau=\SI{2}{s}$ does not contradict the results in \cite{Hong:2007,Olmi:2014} and raises an interesting issue for future investigation as to how the difference between $K_c^{\rm num}$ and $K_c^{\rm theory}$ scales with $N$.
%
However, in cases with supercritical bifurcation, the numerically
observed value of the critical coupling seems more different from the theoretical value than in cases with subcritical bifurcation (which could be another indicator of the type of transition).
We thus have no prior knowledge of the exact value $K_c^{\rm num}(N)$ at which to expect the transition, see right plot in Fig.~\ref{fig:num_results_supp1}. 
Furthermore, the multistability present in the system due to the delayed interaction results in a number of step-like transitions that follow once the incoherent state becomes unstable.
As a consequence,  validation of a linear and continuous transition for $K>K_{\rm c}$ becomes hard to resolve when increasing $K$ further than the regime of the continuous transition, see left plot in Fig.~\ref{fig:num_results_supp1}.
In that case, $\bar{R}$ does not return through the same values since the first bifurcation has already been followed by one or more other bifurcations. 

For smaller values of the inertia, the discontinuous transitions and hysteresis regimes also become much smaller, and it becomes difficult to resolve them even with, e.g., $N=10^5$ oscillators.
Furthermore, there are no a priori conditions to guide 
our choice of the discretization $\Delta K$ with which we change the coupling strength and the time $T_{\rm num}$ for which the coupling strength is kept constant.
The tests required in estimating feasible values for the dynamical parameters are computationally expensive and require long computation times.

Note that the cases in which the predicted  discontinuous transitions appear continuous in numerics~(e.g. Fig.\ref{fig:num_results_m0p1} left panel, third row) are only due to the simulation time being too short in the bifurcation region for the instability to grow enough and yield a large value of the order parameter. 

\section{Conclusions}
\label{conclusions}
In this work, we have studied the effect of time delay in the interaction between oscillators within the framework of the inertial Kuramoto model of globally coupled oscillators.
For a generic choice of the natural frequency distribution of the oscillators, we obtain exact analytical
results that imply that in contrast to the case with no delay, the system in the stationary state may
exhibit either a subcritical or a supercritical bifurcation between a synchronized and an incoherent phase. 
The precise nature of bifurcation has an essential dependence on the amount of delay present in the interaction as also on the value of inertia of the oscillators. 
Our theoretical analysis, performed in the limit of an infinite number of oscillators, is carried out by employing an unstable manifold expansion in the vicinity of the bifurcation, which we apply
to the kinetic equation satisfied by the single-oscillator distribution function, Eq.~(\ref{eq:del:pde}). 
The one-dimensional reduction, Eq.~(\ref{eq:A_red}), of the dynamics for the order parameter is plagued by singularities that are reminiscent of an infinite dimensional bifurcation and, thus, gives at best qualitative information on the bifurcation nature, a fact that our numerical results fully support. We notice, however, that the unstable manifold method is very robust in the context of kinetic equations with continuous spectrum, since it is, to the best of the authors' knowledge, the only one giving analytic predictions for the Kuramoto model both with inertia $m\neq 0$ and delay $\tau\neq 0$, while other methods, like the Ott-Antonsen ansatz~\cite{Ott:2008}, work only for $m=0$, while  self-consistent methods e.g.~\cite{Tanaka:1997} have been applied only for the case with no delay, $\tau=0$.
Direct numerical integration of the dynamics allows to highlight the subtleties one is confronted with when checking the analytical results against those obtained numerically for a finite number of oscillators.  
For systems of delay-coupled PLLs with heterogeneous natural frequencies, our results allow to predict the minimal coupling sensitivity of the voltage-controlled oscillators necessary to enable the network to become synchronized. 
Moreover, such PLL networks generally seem to exit the incoherent state
at smaller coupling sensitivity if the transition happens through a
subcritical bifurcation, $\Re(c_3)>0$, and close to integer multiples of the mean natural period of the oscillators, where we find the local minima of $K_c$, see Fig.~\ref{fig:del:strog_m}.
It may be noted that with increasing transmission delay, the onset of synchronization can generally be achieved at smaller values of $K_c$.
Hence, larger values of the transmission delay seem to decrease the stability of the incoherent state.


\begin{acknowledgements}
DM gratefully
acknowledges the support of the U.S. Department of Energy through the
LANL/LDRD Program and the Center for Non Linear Studies, LANL. S.G. acknowledges support from the Science
and Engineering Research Board (SERB), India under SERB-TARE scheme Grant No.
TAR/2018/000023 and SERB-MATRICS scheme Grant No. MTR/2019/000560. This work was done during SG's visit to the Max Planck Institute
for the Physics of Complex Systems, Dresden, Germany during November 2016 and September
2018 and his visit to the International Centre for Theoretical Physics,
Trieste, Italy (as a Regular Associate of the Quantitative Life Sciences
Section) and Sapienza Universit\`{a} di Roma, Rome, Italy
during May-June 2019. He thanks these organizations as well as his parent organization, Ramakrishna Mission Vivekananda Educational and
Research Institute, for supporting his visits. 
This work was partly supported by the Federal Ministry of Education and Research (BMBF) under the reference number 03VP06431.
\end{acknowledgements}


\end{document}